\documentclass{aa}

\usepackage{float} 

\usepackage{textcomp}

\usepackage{amsmath, amssymb, amsfonts, amscd}
\usepackage{mathtools} 
\usepackage{array} 
\usepackage{cases} 
\usepackage{empheq} 
\usepackage{slashbox} 
\usepackage{multirow} 
\usepackage{txfonts} 

\usepackage{graphicx} 
\usepackage{wrapfig} 

\usepackage{color} 
\usepackage{ulem} 
\usepackage[colorlinks=true,urlcolor=blue]{hyperref} 

\usepackage[toc,page,titletoc]{appendix}

\DeclareSymbolFont{boldsymbols}{OMS}{cmsy}{b}{n}
\DeclareSymbolFontAlphabet{\mathbfcal}{boldsymbols}

\newcommand{\mcal}{\mathcal}

\newcommand{\mbf}{\mathbf}
\newcommand{\mrm}{\mathrm}
\newcommand{\ora}{\overrightarrow}
\newcommand{\I}{{\rm I}}
\newcommand{\II}{{\rm II}}

\renewcommand{\Re}{\mathcal R \!e}
\renewcommand{\Im}{\mathcal I\!m}

\newcommand{\se}{\geqslant}

\newcommand{\intervalle}[3]
 { #1 \in \left\llbracket #2,#3 \right\rrbracket }
 
\newcommand{\interval}[3]
 { #1 \in \left[ #2,#3 \right] }

\begin{document}

\title{The equilibrium tide in stars and giant planets\\ I -- the coplanar case}

\author{F. Remus\inst{1, 2, 3}, S. Mathis\inst{3, 4}, J.-P. Zahn\inst{1}}
\offprints{F. Remus}

\institute{
 LUTH, Observatoire de Paris, CNRS, Universit\'e Paris Diderot, 5 place Jules Janssen, 92190 Meudon, France
 \and
 IMCCE, Observatoire de Paris, UMR 8028 CNRS, UPMC, USTL, 77 avenue Denfert-Rochereau, 75014 Paris, France
 \and
 Laboratoire AIM, CEA/DSM, CNRS, Universit\'e Paris Diderot, IRFU/SAp Centre de Saclay, 91191 Gif-sur-Yvette, France
 \and
 LESIA, Observatoire de Paris, CNRS, UPMC, Universit\'e Paris Diderot, 5 place Jules Janssen, 92190 Meudon, France
 \\{}\\{}
 \email{francoise.remus@obspm.fr, stephane.mathis@cea.fr, jean-paul.zahn@obspm.fr} 
}

\date{Received September 27, 2011; accepted May 9, 2012}

\abstract
{ Since 1995, more than 500 extrasolar planets have been discovered orbiting very close to their parent star, where they experience strong tidal interactions. 
Their orbital evolution depends on the physical mechanisms that cause tidal dissipation, and these are still not well understood. 
}
{
 The purpose of this paper is to refine the theory of the equilibrium tide in fluid bodies that are partly or entirely convective, to predict the dynamical evolution of the systems. In particular, we examine the validity of modeling the tidal dissipation by the quality factor~$Q$, as is commonly done. We consider here the simplest case where the considered star or planet rotates uniformly, all spins are aligned, and  the companion is reduced to a point-mass. 
}
{ We  expand the tidal potential in Fourier series, and express the hydrodynamical equations in the reference frame, which rotates with the corresponding Fourier component. The results are cast in the form of a complex disturbing function, which may be implemented directly in the equations governing the dynamical evolution of the system.}
{The first manifestation of the tide is to distort the shape of the star or planet adiabatically along the line of centers. This generates the divergence-free velocity field of the adiabatic equilibrium tide, which is stationary in the frame rotating with the considered Fourier component of the tidal potential; this large scale velocity field is decoupled from the dynamical tide. The tidal kinetic energy is dissipated into heat through turbulent friction, which is modeled here as an eddy-viscosity acting on the adiabatic tidal flow. This dissipation induces a second velocity field, the dissipative equilibrium tide, which is in quadrature with the exciting potential; it is responsible for the imaginary part of the disturbing function, which is implemented in the dynamical evolution equations, from which one  derives characteristic evolution times.
}
{\rm The rate at which the system evolves depends on the physical properties of tidal dissipation, and specifically on how the eddy viscosity varies with tidal frequency and on the thickness of the convective envelope for the fluid equilibrium tide. At low frequency, this tide retards by a constant time delay, whereas it lags by a constant angle when the tidal frequency exceeds the convective turnover rate.}


\titlerunning{The equilibrium tide in stars and giant planets -- The coplanar case}
\authorrunning{ F. Remus et al.}

\maketitle

\section{Introduction and general context}
\label{intro}

Stellar binaries provide precious information on the structure of stars and the dynamical processes operating in them. 
Analysis of the motion of a binary's components allows to determine their masses and radii, and through the apsidal precession it puts constraints on their internal mass distribution.
Furthermore, their orbital evolution due to tidal dissipation sheds some light on their formation and past history. 
Quite naturally, similar benefits may be expected from the study of extrasolar planets. 
Indeed, since 1995,  more than five hundred of those have been discovered  orbiting very close to their host star, and their number keeps increasing thanks to dedicated space missions (CoRoT and KEPLER) and vast ground-based programs, with powerful instruments such as HARPS { (for a review, see Udry \& Santos 2007 and references therein)}. 
Furthermore, recent studies on the orbital evolution of natural satellites in our solar system have allowed an improved quantification of the tidal dissipation in Jupiter ({Lainey et al. 2009}) and Saturn ({Lainey et al. 2012}). 
This renewed interest for the gravitational interaction between various celestial bodies motivates us to refine our knowledge of the physical processes that intervene in tides.\\

The dynamical evolution of a binary system is driven by the conversion of its mechanical energy into heat. 
Provided the system loses no angular momentum, it tends to the state of minimum energy in which the orbits are circular, the rotation of the components is synchronized with the orbital motion, and the spins are aligned. 
However, in very close systems with high mass ratio, such final state cannot be achieved: instead, the planet spirals toward the star and may eventually be engulfed by it (Hut 1981; Levrard et al. 2009). 
To predict the fate of a binary system, one has to identify the dissipative processes that achieve the conversion of kinetic energy into thermal energy, from which one may then draw the characteristic times of circularization, synchronization and spin alignment. 
Before reviewing these processes, let us recall the two types of tides which operate in stars and in the fluid parts of giant planets. 
The {\it equilibrium tide} designates the large-scale flow induced by the hydrostatic adjustment of the star in response to the gravitational force exerted by the companion { (Zahn 1966; we shall refer to that paper as Z66)}. 
On the other hand, the  {\it dynamical tide} corresponds to the eigenmodes (gravity, inertial, or gravito-inertial waves) that are excited by the tidal potential { (Zahn 1975; Goldreich \& Nicholson 1989; Ogilvie \& Lin 2004; Goodman \& Lackner 2009; Rieutord \& Valdetarro 2010; Barker \& Ogilvie 2010)}. 
These tides experience two main dissipative mechanisms: turbulent friction in convective regions and thermal dissipation acting on the gravito-inertial modes excited in radiative zones.\\

We shall focus here on the equilibrium tide acting in the convective envelopes of solar-like stars or giant planets. 
The physical description of this tide, with turbulent dissipation, has been formulated by {Z66}. 
Several authors have applied this theory to binary stars (see Koch \& Hrivnak 1981; Bouchet \& Zahn 1989; Verbunt \& Phinney 1995), and they found that it agreed well with the observations  (pre-main-sequence binaries, red giants, etc.), except for main-sequence stars older than about one Gyr (Mathieu et al. 2004). 
Here, we revisit the theoretical formalism of Z66 by expressing the velocity field of the equilibrium tide in the appropriate reference frame, i.e. that which rotates with each Fourier component of the tidal potential. 
This treatment filters out properly the dynamical tide, suppressing the spurious `pseudo-resonances' encountered in Z66. 
Our second goal, in the present work, is to improve the description of the equilibrium tide by tying the quality factor $Q$, which is commonly used in planetology and celestial mechanics, with the physical process of {convective} turbulent dissipation (Goldreich \& Soter 1966, Goldreich \& Nicholson 1977). 
We shall assume here, for simplicity, that all spins are aligned, i.e. that the rotation axes of star or planet are perpendicular to the orbital plane; the case with non-aligned spins  will be treated in a forthcoming paper. 
{Furthermore, the differential rotation, which is present in stellar and planetary interiors is not accounted for in this first work and the rotation is taken as uniform.}



\section{Reference frame and tidal potential}


\subsection{The reference frame}
We consider a system consisting of two bodies A and B, of mass $m_A$ and $m_B$, and we undertake to describe the tide exerted by B on A, which we assume to be in fluid state, i.e. a star or a giant planet. 
Due to their mutual attraction, they move in elliptic orbits around their common center of mass, but it is often convenient to choose a reference frame $\mcal{R}_A$ whose origin is placed at the center of A (designated by the same letter) and whose axes {$\{\bf X, \bf Y, \bf Z \}$} have fixed directions in space. 
We assume that all spins are aligned along the $\left(A{\rm\bf Z}\right)$ axis; therefore the position of B is entirely determined by the three following keplerian elements~: the semi-major axis $a$ of the relative orbit of B around A, its eccentricity $e$, and its mean anomaly $M = n t$ where $n$ is the mean motion.

\vspace{0.5cm}
\begin{figure}[!htb]
 \centering
 \includegraphics[width=\linewidth] {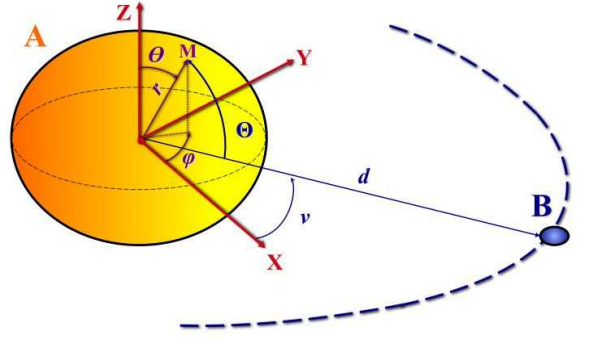}
 \caption{
 Spherical coordinates system attached to the reference frame $\mcal{R}_A$ centered on A. 
 The spin of Ais perpendicular to the orbital plane of B. 
 The dashed line illustrates the orbit of B.
 \label{Fig_Repere}
 }
\end{figure}
\vspace{0.5cm}


\subsection{The tidal potential}


 The tidal force exerted by B on A derives from a potential, which may be expanded in Legendre polynomials as follows :
\begin{equation}
{U_E}\left(\vec r,t\right) = -\sum_{s \se 2} \; \frac{\mcal{G} m_B}{d} \, 
 \left( \frac{r}{d} \right)^s \, \mrm{P}_s \left( \cos\Theta \right) \;,
 \label{UE}
\end{equation}
where $\Theta = ( \ora{AB} , \ora{AM)}$, $M$ being the current point of position $\vec r$, and $d$ the distance $AB$.
Kepler's third law $\displaystyle \mcal G \, (m_A +m_B)=  n^2 \, a^3$ allows us to rewrite (\ref{UE}) in a slightly different form:
\begin{equation}
{U_E}\left(\vec r,t\right) = -\frac{m_B}{m_A+m_B} \,  n^2 \sum_{s \se 2} \; a^3
 \, \frac{r^s}{ d^{s+1}} \, \mrm{P}_s \left( \cos \Theta \right) \;.
\end{equation}
Finally, introducing the angular variables $(\theta, \varphi)$ related to the reference frame $\mcal{R}_A$ and the associated Legendre functions $ \mrm{P}_s^m \left(\cos\theta\right) $, we get 
\begin{multline}
 \label{UE_Legendre}
{U_E}\left(r,\theta,\varphi,t\right) = -\frac{m_B}{m_A+m_B}  \, n^2 \sum_{s \se 2} \, \Re \left[
 \frac{r^s}{a^{s-2}} \, \sum_{m=0}^{s} \, \frac{(s-m)!}{(s+m)!}
 \right. \\
 \left. \times\, \left(2-\delta_{0,m}\right) \mrm{P}_s^m \left(0\right) \, \mrm{P}_s^m \left(\cos\theta\right) \, \exp \left[ \pmb i m \varphi \right]
 \left( \frac{a}{d} \right)^{s+1}\, \exp \left[-\pmb i m v\right] \right] \: ,
\end{multline}
where $v$ designates the true anomaly of B. 
The next step is to expand the time-dependent terms in Fourier series:
\begin{equation}
 \left( \frac{a}{d} \right)^{s+1}\,  \exp \left [-\pmb i m v \right]
 = \sum_{l} G_{s,\frac{s-m}{2},l-m} \left(e\right) \,  \exp \left[-\pmb i l n t \right] \: ,
\end{equation}
where we made use of Kaula's functions $G_{j,p,q}(e)$ (Kaula 1962).  Furthermore, we have:
\begin{equation}
 \mrm P_s^m\left(0\right) = \begin{cases}
    (-1)^{\frac{s-m}{2}} \, \dfrac{\left(m+s\right)!}
       {2^s \left( \frac{m+s}{2} \right)! \,\left( \frac{m-s}{2} \right)! }
       & \text{, if $\left(s+m\right)$ is even} \\
    0 & \text{, if $\left(s+m\right)$ is odd}
                   \end{cases} \,.
\end{equation}

We chose the Kaula's functions in the Fourier expansion, instead of the Hansen coefficients, to prepare the ground for a subsequent paper where we shall treat the case of non-aligned spins. 
These functions $G_{s,p,q}\left(e\right)$ are polynomials in $e$ whose lowest power is ${|q|}$. 
We give their expressions in Table \ref{Gspqe} for the most frequently used triplets $\left\{s,p,q\right\}$, since for small eccentricity it will suffices to retain only a few  $q$ values ($\intervalle{q}{-2}{2}$).

From now on, we shall keep only the time-dependent part of the potential $U_E$, which we call $U$, the tidal potential:
\begin{equation}
   U \left(r,\theta,\varphi,t\right)  =  \sum_{s \se 2} \sum_{\underset{\left(s+m\right)\;{\rm even}}{m \se 1}}^s \sum_l U_s^m \left(r\right)\, \mrm{P}_s^m \left(\cos  \theta\right) \, \cos \left(m\varphi - ln t\right)\,,  
 \label{U}
\end{equation}
where the radial parts $U_s^m$ of the potential are given by:
\begin{equation}
  U_s^m(r)  = - 2 \frac{m_B}{m_A + m_B} \, n^2
                 \, \frac{(s-m)!}{2^s \left( \frac{s+m}{2} \right)! \left(\frac{s-m}{2}\right)!}
                 \, \frac{r^s}{a^{s-2}} \, G_{s, \frac{s-m}{2},l-m}\left(e\right)\, ;
\end{equation}
{the tidal potential is thus the interaction potential, from which we have subtracted the acceleration responsible for the movement of the center of mass of A: $\mathcal G\,m_B/d\,\left[1+\left(\vec r_B\cdot\vec r\right)/d^2\right]$.}

\begin{table}[H]
 \centering
 \begin{tabular}{c c c c c c c l l}
  \hline
  \hline
  \multicolumn{6}{c}{} & \vline &\multicolumn{1}{c}{}\\
  $s$ & $p$ & $q$ & $s$ &  $p$ & $q$ & \vline & $G_{s,p,q}\left(e\right)$ \\
  \multicolumn{6}{c}{} & \vline &\multicolumn{1}{c}{}\\
  \hline
  & & & & & & \vline & \\
  2 & 0 & -2 & 2 & 2 &  2 & \vline & 0\\
  & & & & & & \vline &\\
  2 & 0 & -1 & 2 & 2 & 1 & \vline & $-\frac{1}{2}e+\dots$\\
  & & & & & & \vline & \\
  2 & 0 & 0 & 2 & 2 &  0 & \vline & $1-\frac{5}{2}e^{2}+\dots$\\
  & & & & & & \vline &\\
  2 & 0 & 1 & 2 & 2 & -1 & \vline & $\frac{7}{2}e+\dots$\\
  & & & & & & \vline &\\
  2 & 0 & 2 & 2 & 2 & -2 & \vline & $\frac{17}{2}e^{2}+\dots$\\
  & & & & & & \vline &\\
  2 & 1 & -2 & 2 & 1 & 2 & \vline & $\frac{9}{4}e^{2}+\dots$\\
  & & & & & & \vline &\\
  2 & 1 & -1 & 2 & 1 & 1 & \vline & $\frac{3}{2}e+\dots$\\
  & & & & & & \vline &\\
  &  &  & 2 & 1 & 0 & \vline & $\left(1-e^{2}\right)^{-3/2}$\\
  & & & & & & \vline &\\
  \hline
  \hline
 \end{tabular}
 \caption{Values of the eccentricity function $G_{s,p,q}\left(e\right)$ in the quadrupolar case where $s=2$. Note their symmetry property $G_{s,p,q}(e) = G_{s,s-p,-q}(e)$. (From Lambeck 1980)}
\label{Gspqe}
\end{table}

We may further simplify this expression of the tidal potential if we consider binary systems that are separated enough to allow the companion to be treated as a point mass; this approximation has been discussed by Mathis \& Le Poncin-Lafitte (2009). 
In this {\it quadrupolar approximation}, we shall retain only the lowest term of the potential ($s=2$). Furthermore, since we are considering the coplanar case, the sole value of $m$ that still remains is $m=2$ in Eq. (\ref{U})  and $U$ thus becomes
\begin{equation}
U = - \frac{1}{4} \frac{m_B}{m_A + m_B} n^2 r^2 \sum_l  G_{2,0,l-2}(e) P_2^2(\cos \theta)
\cos (2 \varphi - ln t) .  \label{U2}
\end{equation}

\section{The tidal velocity field}

\subsection{Governing equations}

The tidal potential (\ref{U2}) induces in the star (or the giant planet) pressure and density perturbations and a modified velocity field $\mbf W$, which obeys the following equations of motion:
\begin{equation}
\rho \, \frac{\mrm D \mbf W}{\mrm Dt}
   = -\nabla  P - \rho \, \nabla \left(\Phi+U\right) + \mbf F_{\rm visc} \, , \label{eqNS}
\end{equation}
of continuity:
\begin{equation}
 \frac{\mrm D \rho}{\mrm Dt} + \rho \nabla \cdot \mbf W = 0\, , \label{eqCONT}
\end{equation}
Poisson:
\begin{equation}
  \nabla^2 \Phi = 4 \pi \mcal{G} \rho \, , \label{eqPOIS}
\end{equation}
and of entropy:
\begin{subequations}
 \begin{empheq}
 [left=\displaystyle{ \rho \frac{\mrm D S}{\mrm Dt} = \empheqlbrace \,} , right=\text{,}]
 {flalign}
   &\frac{\rho \varepsilon - \nabla \cdot \mbf H}{ T}
      &&\text{(for a radiation zone)}     &&&&&& \label{Srad}\\
   &\mrm{cste} &&\text{(in a convective zone)} &&&&&& \label{Sconv}
 \end{empheq}
\end{subequations}
where we made use of the total derivative
\begin{equation}
\frac{\mrm D }{\mrm Dt} =
      \frac{\partial} {\partial t} + \left( \mbf W \cdot \nabla \right) \: .
\end{equation}      
These are completed by the expressions for the radiation flux:
\begin{equation}
  \mbf H = -\frac{16 \sigma}{3\kappa\rho} \,  T^3 \, \nabla  T \, ,
\end{equation}
and for the specific entropy:
\begin{equation}
   S =\frac{\mcal R_{g}}{(\gamma-1)\mu} \ln \left( \frac{ P}{\rho^\gamma} \right) \,.
\end{equation}

The physical quantities, which intervene in these equations, are the pressure $ P$, the temperature $ T$, the density $\rho$, the gravitational potential created by the star $\Phi$, the viscous force $\mbf F_{\rm visc}$, the nuclear energy production rate by unit mass $\varepsilon$, the absorption coefficient $\kappa$, the ratio $\gamma={C_{ P}}/{C_{ V}}$ of specific heats at constant pressure ($C_{ P}$) and constant density ($C_{ \rho}$), the molecular mass $\mu$. $\sigma$ is the Stefan-Boltzmann constant, and $\mcal R_{g}$ the perfect gas constant. \\{}

Treating the tides as a small amplitude perturbation of the hydrostatic structure of the object, we split all scalar quantities as
\begin{equation}
   X\left(r,\theta,\varphi,t\right) = X_0\left(r\right) + X' \left(r,\theta,\varphi,t\right) \, ;
\label {split}
\end{equation}
$X_0$ designates the spherically symmetrical profile of $X$, and $X'$ represents the perturbation due to the tidal potential. 
The reference state is drawn from an up-to-date stellar or planetary structure model.  

We now turn to the global velocity field $  \mbf W$, from which we subtract the star's rotation (assumed here to be uniform) to isolate the time dependent tidal velocity field $\mbf V'$: 
\begin{equation*}
  \mbf W \left(r,\theta,\varphi,t\right) = \mbf\Omega \wedge \mbf r + \mbf V'\left(r,\theta,\varphi,t\right)  \,.
\end{equation*}
\\{}
This tidal velocity field may be expanded in the same way as the tidal potential (\ref{U}), and each component $\mbf V'_{m,l}$ may be 
analyzed in the reference frame ${\mcal R_C}$ rotating with the corresponding tidal frequency, i.e. the angular velocity
\begin{equation}
\Omega^R = \frac{l}{2} \, n\, .
\end{equation}
In that reference frame, all quantities $X'$ induced by the equilibrium tide are stationary, in particular the density and the velocity field: 
\begin{equation}
\label{defRC}
  \left. \frac{\mrm D \rho'_{m,l} }{\mrm Dt} \right|_{\mcal R_C}= 0 \,, \qquad
  \left. \frac{\mrm D \mbf V'_{m,l} }{\mrm Dt} \right|_{\mcal R_C}= \mbf 0 \,.
\end{equation}
From the continuity equation (\ref{eqCONT}), we then deduce that 
\begin{equation}
  \nabla \cdot \mbf V'_{m,l} \, = \, 0 \, ;
\end{equation}
this property will hold in any reference frame, and thus for the total tidal velocity field $\mbf V'$.
This confirms the solenoidal character of that flow, contrary to the claim made by Scharlemann (1981).



\subsection{Adiabatic system~-~Dissipative system}
\label{Ad_Dissip}

In our treatment of the equilibrium tide, we shall separate all quantities and equations in two parts. \\{}
\begin{itemize}
\item
First, an {\it adiabatic system} (I) in phase with the perturbing potential $U$ (eq. \ref{U2})   : it corresponds to the adiabatic tide, i.e. the star's response to the tidal excitation, ignoring  all dissipative processes. \\{}

\item
Second, a {\it dissipative system} (II) in quadrature with the perturbing potential~: it corresponds to  the star's response due to the dissipative processes (here, the turbulent friction due to the convective motions).
\end{itemize}

\begin{figure}[!htb]
\begin{center}
	\includegraphics[width=\linewidth]{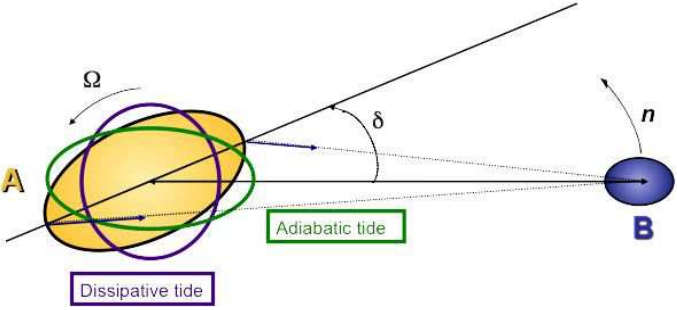}
	\vskip 15pt
	\caption{
Typical system of tidal interaction. 
{\rm Body B exerts a tidal force on body A, which adjusts itself with a phase lag $\delta$, because of internal friction. 
This adjustment may be split on an adiabatic component, which is in phase with the tide, and a dissipative one, which is in quadrature.}
\label{figMarees}
}
\end{center}
\end{figure}

The scalar perturbations \eqref{split} will thus be split as:
\begin{equation}
X' \left(r,\theta,\varphi,t\right) = X_\I\left(r,\theta,\varphi,t\right) + X_\II\left(r,\theta,\varphi,t\right) \,, 
\end{equation}
where $X_\I$ designates the contribution of the adiabatic tide, in phase with the potential $U$:
\begin{eqnarray}
\lefteqn{X_\I = \sum_{l} {\widetilde X}_{\I,l} (r,\theta) \, \cos\left(2\varphi-ln t\right) \,} \nonumber\\
&=& \sum_{l} {\widehat X}_{\I,l} (r)\,{\rm P}_{2}^{2}\left(\cos\theta\right)\,\cos\left(2\varphi-ln t\right)\,,
\label{XI}
\end{eqnarray}
and $X_\II$ that of the dissipative tide, in quadrature with $ U$:
\begin{eqnarray}
\lefteqn{X_\II = \sum_{l} {\widetilde X}_{\II,l} \left(r,\theta\right) \, \sin\left(2\varphi-ln t\right) \,}\nonumber\\
&=& \sum_{l} {\widehat X}_{\II,l} \left(r\right)\,{\rm P}_{2}^{2}\left(\cos\theta\right)\,\sin\left(2\varphi-ln t\right)\,.
\label{XII}
\end{eqnarray}
We apply the same distinction to the adiabatic tidal flow:
\begin{equation}
  \mbf V' = \mbf V_\I\left(r,\theta,\varphi,t\right) + \mbf V_\II\left(r,\theta,\varphi,t\right)\,,
\end{equation}
where the tidal field $\mbf V_\I$ induces a radial displacement which is in phase with the potential $U$:
\begin{equation}
 \mbf V_\I =  \sum_{l} \left[ 
 \begin{array}{l}
   u_\I(r,\theta) \, \sin(2\varphi-ln t)\\
   v_\I(r,\theta) \, \sin(2\varphi-ln t)\\
   w_\I(r,\theta) \, \cos(2\varphi-ln t)
 \end{array} \right] \,,
\end{equation}
and the velocity field $\mbf V_\II$ associated with the dissipative tide which is in quadrature:
\begin{equation}
 \mbf V_\II =  \sum_{l} \left[ 
 \begin{array}{l}
   \quad\, u_\II(r,\theta) \, \cos(2\varphi-ln t) \\
   \quad\, v_\II(r,\theta) \, \cos(2\varphi-ln t) \\
   -\,     w_\II(r,\theta) \, \sin(2\varphi-ln t)
 \end{array} \right] \,.
\end{equation}

We now linearize the original system (\ref{eqNS} - \ref{eqPOIS}),  and split it likewise in systems (I) and (II), both with their equations of motion, of continuity, Poisson, and of entropy, in this order. 

\paragraph{Adiabatic system (I)}

\begin{equation}
 \rho_0 \left[ \frac{\mrm D \mbf V_\I}{\mrm Dt}
                + 2\mbf\Omega\wedge\mbf V_\I \right] 
   = -\nabla  P_\I - \rho_0 \, \nabla \left( \Phi_\I +  U \right)
     - \rho_\I \, \nabla \Phi_0  \; ,
 \label{eqNS_I}
\end{equation}

\begin{equation}
 \frac{\mrm D \rho_\I}{\mrm D t} + \rho \nabla \cdot \mbf V_\I = 0  \,, 
 \label{eqCONT_I}
\end{equation}

\begin{equation}
\nabla^2\Phi_\I = 4\pi \mcal{G} \rho_\I \,,
 \label{eqPOIS_I}
\end{equation}
 \vspace{-0.3cm}
\begin{equation}
  \rho_0  \frac{\mrm D S_\I}{\mrm D t} = 0 \,. \label{Sconv_I}
\end{equation}

\paragraph{Dissipative system (II)}

\begin{equation}
 \rho_0 \left[ \frac{\mrm D \mbf V_\II}{\mrm Dt} + 2\mbf\Omega\wedge\mbf V_\II \right]
   = -\nabla  P_\II
     - \rho_0 \, \nabla\Phi_\II - \rho_\II \, \nabla\Phi_0 +\mbf F_\II \,,
 \label{eqNS_II}
\end{equation}

\begin{equation}
  \frac{\mrm D \rho_\II}{\mrm D t} + \rho \nabla \cdot \mbf V_\II = 0 \,,
 \label{eqCONT_II}
\end{equation}

\begin{equation}
\nabla^2\Phi_\II = 4\pi \mcal{G} \rho_\II \,,
 \label{eqPOIS_II}
\end{equation}
 \vspace{-0.3cm}
\begin{equation}
  \rho_0   \frac{\mrm D S_\II}{\mrm D t} = 0. \label{Sconv_I}
\end{equation}
As was shown in Z66, the dissipative tide is much weaker than the adiabatic tide.
That means that for a given scalar field $X' = X_\I + X_\II$, we have $ X_\II  \ll X_\I$.
Moreover, this allows us to neglect the viscous force $\mbf F_\I ({\mbf V}_\II)$ in system (I), but we keep $\mbf F_\II ({\mbf V}_\I)$ in system (II), where it induces the dissipative tide.

\subsection{The adiabatic equilibrium tide}

\subsubsection{Adiabatic response to the tidal potential - Love number}

We begin with the study of the adiabatic equilibrium tide, which is the hydrostatic response to the perturbing potential $U$ exerted by the companion body, in the absence of dissipation. We expand the perturbations of pressure as in (\ref{XI}):
\begin{equation}
P_\I\left(r, \theta, \varphi, t\right)= \sum_{l} \widehat{P}_{\I,l}\left(r\right){\mrm P_2^2}\left(\cos\theta\right)\cos\left(2 \varphi - l n t\right) , \\
\end{equation}
and likewise for $\rho_\I(r, \theta, \varphi, t)$ and $\Phi_\I(r, \theta, \varphi, t)$.
These perturbations obey the equation of motion  (\ref{eqNS_I}) of system (I), from which we remove the acceleration term on the l.h.s.:
\begin{equation}
\nabla  P_\I = - \rho_0 \, \nabla \left( \Phi_\I +  U \right)
     - \rho_\I \, \nabla \Phi_0  \; .
\end{equation}  
From the vertical balance we draw
\begin{equation}
{\partial  P_\I \over \partial r} = - \rho_0 \, {\partial \over \partial r}  \left( \Phi_\I +  U \right)
     - \rho_\I \, g_0  \; 
\end{equation} 
and likewise from the horizontal balance, 
\begin{equation}
P_\I = - \rho_0 \, \left( \Phi_\I +  U \right)  \; ,
\end{equation}  
where $g_0(r)={\mcal G m(r)}/{r^2}$ designates the unperturbed gravity.  
Eliminating $P_\I$ we get the following expression for the density perturbation
\begin{equation}
 \rho_\I = -\frac{1}{g_0} \, \frac{{\rm d}\rho_0}{{\rm d}r} \, \left(\Phi_\I +  U\right) \,;
\end{equation}     
we implement it in the perturbed Poisson equation (\ref{eqPOIS_I}), taking into account that $U$ is an harmonic potential (i.e. $\nabla^2 U= 0$), to finally obtain (cf. Sweet 1950):
\begin{equation}
\label{Poisson_phi+U}
  \nabla^2 \left(\Phi_\I +  U\right) - \frac{4\pi \mcal{G}}{g_0}
     \, \frac{\mrm d \rho_0}{\mrm d r}   \, \left(\Phi_\I +  U\right)     = 0 \; .
\end{equation}

The potential perturbation $\Phi_\I$ has the same dependence as $U$ in $(\theta, \varphi, t)$; therefore we may introduce a function $h(r)$ which depends only on the radial coordinate, such that  $(\Phi_\I +  U)= hU$. It obeys the following differential equation:
\begin{equation}
  \frac{ \mrm d^2 h }{ \mrm d x^2 } + \frac{6}{x} \,
   \frac{ \mrm d h }{ \mrm d x} + f(x) \, h = 0 \,, 
 \label{eqh}  
\end{equation}    
where $x=r/R$, $R$ being the total radius. The function $f(x)$ depends on the internal mass distribution at equilibrium:
\begin{equation}
 f(x) = -\frac{4\pi \mcal G  R}{g_0} \, \frac{\mrm d \rho_0}{\mrm d x} \, .
 \label{f}
\end{equation}

The solution of  (\ref{eqh}) is constrained by two boundaries conditions: regularity at center $x=0$, and $h \rightarrow 1$ as $x \rightarrow \infty$; it may be written as
\begin{eqnarray}
  \Phi_\I(x,\theta,\varphi,t) &=&
   \sum_{l} {\widehat \Phi}_{\I,l} (x)\,{\rm P}_{2}^{2}\left(\cos\theta\right)\,\cos\left(2\varphi-ln t\right)  \nonumber \\
   & =&  \left[ h(x) - 1 \right] \,  U (x,\theta,\varphi,t) \\
   & =& \left[ h(x) - 1 \right] \,
    \sum_{l}  U_l \, x^2 \, P_2^2(\cos \theta) \cos (2\varphi -ln t) \, ,  \nonumber
    \label{PhiI_hs}
\end{eqnarray}
where from (\ref{U2})
\begin{equation}
U _l= - \frac{1}{4} \frac{m_B}{m_A + m_B} n^2 R^2 \;  G_{2,0,l-2}(e)  .  \label{U2bis}
\end{equation} 
The surface value of the ratio $\Phi_\I / U$ is the second-order Love number in the adiabatic case:
\begin{equation}
	k_2^\mrm{ad} = \frac{ \widehat{\Phi}_{\I,l}\left(1\right) }{ U_{l} } 
	         = h\left(1\right)-1.
\label{k2def}
\end{equation}

This constant is twice the classical apsidal motion constant, and it is approximatively the fifth moment of the mass distribution (see Eq. 2.15 of Zahn, 1966a).

\subsubsection{The poloidal component of the adiabatic velocity field}

We have established earlier that each Fourier component of the tidal perturbation is stationary in the reference frame ${\mcal R_C}$ rotating with the angular velocity $\Omega^{R}=(l/2) \, n$. This is true in particular for the potential $(\Phi_\I + U)$, and therefore 
\begin{equation}
\label{VIr}
  \frac{\mrm D}{\mrm Dt}(\Phi_\I+ U)+\mbf V_\I \cdot \nabla \Phi_0 = 0 \; .
\end{equation}
From this we draw the vertical component of the velocity field
\begin{equation}
   V_{\I,r} 
 = \frac{h x^2}{g_0} \sum_{l} \sigma_{l}\left(n,\Omega\right)
   \, U_l  \, P_2^2(\cos \theta) \sin(2\varphi - ln t)\,, 
 \label{Vr2}
\end{equation}
where its amplitude varies linearly with the tidal frequency 
\begin{equation}
\sigma_l = ln - 2\Omega\,. 
\end{equation}

We have now to complete this vertical field with its horizontal counterpart. Since $\mbf V_\I$ is divergence free, with a radial component, it is natural to seek it in form of a poloidal field.
Following Z66 and Rieutord (1987), we write 
\begin{equation}
  \mbf V_\I^{P} \, {\equiv} \,
        \nabla \wedge \nabla \wedge \left[ \left(r\chi^{P}\right) \mbf r \right] \,,
\end{equation}
and expand the poloidal current function $\chi^{P}\left(x, \theta, \varphi, t\right)$ in its Fourier components:
\begin{equation}
    \chi^P\left(x,\theta,\varphi, t\right)
       = \sum_l \xi^P_l\left(x\right) \, \mrm P_2^2\left(\cos\theta\right) \, \sin \left(2\varphi - ln t\right) \, .
       \label{defxiP}
\end{equation}

In the inertial frame the poloidal velocity field is thus given by
\begin{align}
  \mbf V_\I^P
    &= \sum_l \left[ \begin{array}{c}
          \\
          6 \, \xi_l^P \, \mrm P_2^2\left(\cos\theta\right) \, \sin\left(2\varphi - ln t\right)\\ \\
         \dfrac{1}{x} \dfrac{d}{dx} \left(x^2  \xi_l^P \right)\, \dfrac{\mrm d \mrm P_2^2\left(\cos\theta\right)}{\mrm d \theta}
                                              \, \sin\left(2\varphi - ln t\right)\\  \\
         \dfrac{2}{x} \dfrac{d}{dx}  \left(x^2  \xi_l^P \right)\, \, \dfrac{\mrm P_2^2\left(\cos\theta\right)}{\sin\theta} \, \cos\left(2\varphi - ln t\right)
          \\{}
      \end{array} \right]  \,. \label{VIP_hs}
\end{align}
Comparing the radial component with (\ref{Vr2}) we obtain the following expression for $ \xi_l^P(x)$:
\begin{equation}
 \xi_l^P = \frac{1}{6} \, \sigma_l \, \frac{h x^2}{g_0}
       \, U_l  \, . \label{chif}
\end{equation}

\subsubsection{The toroidal component of the adiabatic velocity field}

The description of the adiabatic velocity field is not yet complete. 
Indeed, in the presence of rotation, the equation of motion (\ref{eqNS_I}) can not be satisfied by the sole poloidal velocity field, but one must add a toroidal field
\begin{equation}
  \mbf V_\I^T {=}
        \nabla \wedge \left( \chi^T \mbf r \right) \,,
\end{equation}
where $\chi^T$ is the  toroidal stream function. Like $\chi^P$ before, this function may be expanded in Fourier series:
\begin{equation}
    \chi^T(x,\theta,\varphi, t)
       = \sum_l \chi^T_l\left(x, \theta\right) \, \cos \left(2\varphi - ln t\right) \, , \label{defxiT}
\end{equation}
with the following projections in spherical coordinates
\begin{align}
  \mbf V_\I^T
    &= \sum_l \left[ \begin{array}{c}
          \\
          0 \\ \\
          -  \dfrac{2}{\sin\theta} \, \chi^T_l \, \sin \left(2\varphi - ln t\right) \\  \\
         -  \dfrac{\partial\chi^T_l} {\partial \theta} \, \cos\left(2\varphi - ln t\right)
          \\{}
      \end{array} \right]  \,. \label{VIT_hs}
\end{align}

Here, our treatment departs from that followed in Z66. Instead of working in the inertial frame, we adopt the reference frame ${\mcal R_C}$ rotating with the Fourier component of the tidal potential potential $U$, i.e. with the angular velocity $\Omega^R = \frac{l}{2} \, n$. In this frame, all quantities are stationary; hence, the momentum equation reduces to
\begin{equation}
 \rho_0 \, 2 \mbf\Omega^R\wedge\mbf V_\I 
   = -\nabla  P_\I - \rho_0 \, \nabla \left( \Phi_\I +  U \right)
     - \rho_\I \, \nabla \Phi_0  \; ,
 \label{eqNS_Ix}
\end{equation}
for each Fourier component.
Their $\theta$ and $\varphi$ projections may be written
\begin{subequations}
 \begin{equation}
   2 \Omega^R \, \cos\theta \left[ \, \frac{\partial \chi_l^T}{\partial\theta} 
   -   \frac{\mrm P_2^2\left(\cos\theta\right)}{\sin\theta} \, {2\over x} {d \over dx}(x^2 \xi^P_l )\,\right]
      = \frac{\partial \mcal F\left(r,\theta\right)}{\partial\theta} \,,
 \end{equation}
and
 \begin{multline}
  2 \Omega^R \left[ \, \frac{\cos\theta}{\sin \theta}  \, 2 \,\chi_l^T
  + \, \sin\theta \, \mrm P_2^2\left(\cos\theta\right)\, 6 \, \xi^P_l\,\right. \\
  + \left. \,  
      \cos\theta \frac{\mrm d \mrm P_2^2\left(\cos\theta\right)}{\mrm d\theta} {1 \over x} {d \over dx}(x^2 \xi^P_l ) \, \right]
   = \frac{2 \mcal F\left(r,\theta\right)}{\sin\theta} \,, 
  \end{multline}
\end{subequations}
where the function $\mcal F$ includes all functions of the r.h.s. of that momentum equation. Its elimination, equivalent of taking the curl of (\ref{eqNS_Ix}), leads to
\begin{equation}
\label{eqchil}
\begin{split}
 2 \, \chi_l^T
  =  &\quad 6 \, \xi_l^P \, \frac{1}{\sin\theta} \, \frac{\mrm d}{\mrm d\theta}
                                  \left[ \sin^2\theta \, \mrm P_2^2\left(\cos\theta\right) \right]
 \\
   &- {1 \over x} {d \over dx}\left(x^2 \xi^P_l\right) \left[ 6 \, \cos\theta \,\mrm P_2^2(\cos\theta)
  + \sin\theta \, \frac{\mrm d \mrm P_2^2}{\mrm d\theta} \right] \, .
\end{split}
\end{equation}

After some straightforward manipulations (see Z66), we get 
\begin{equation}
 \label{cl2}
  \chi^T = \sum_{l} \xi_l^T \left(x\right) \mrm P_3^2 \left(\cos\theta\right) \, \cos\left(2\varphi-ln t\right) \; ,
\end{equation} 
with
\begin{equation}
\quad \xi_l^T =
 \label{xiT}
   \frac{2}{15} \, \sigma_l \, x^2
 \frac{\mrm d}{\mrm dx} \left( \frac{hx}{g_0} \right) \, U_l \, ,
\end{equation}
and reach finally the following expression for the toroidal field, in the inertial frame:
\begin{multline}
    \mbf V_\I^T = - \sum_{l} \frac{2}{15}  \, \sigma_l \, x^2
        \, \frac{\mrm d}{\mrm dx} \left( \frac{hx}{g_0} \right) \, U_l \\
      \times \, \left[ \begin{array}{c}
          \\
          0 \\ \\
          2 \dfrac{\mrm P_3^2(\cos\theta)}{\sin\theta} \, \sin\left(2\varphi-ln t\right) \\ \\
          \dfrac{\mrm d \mrm P_3^2(\cos\theta)}{\mrm d\theta} \, \cos \left(2\varphi-ln t\right)
          \\{}
      \end{array} \right]  \,.
\end{multline}

Using the reference frame rotating with each component of the tidal potential, we are thus able to decouple rigourously the equilibrium tide, which is stationary in this frame, and the dynamical tide. Therefore, the `pseudo-resonances' found in Z66 disappear.

\begin{figure*}[!htp]
\begin{center}
 \includegraphics[width=0.375\textwidth]{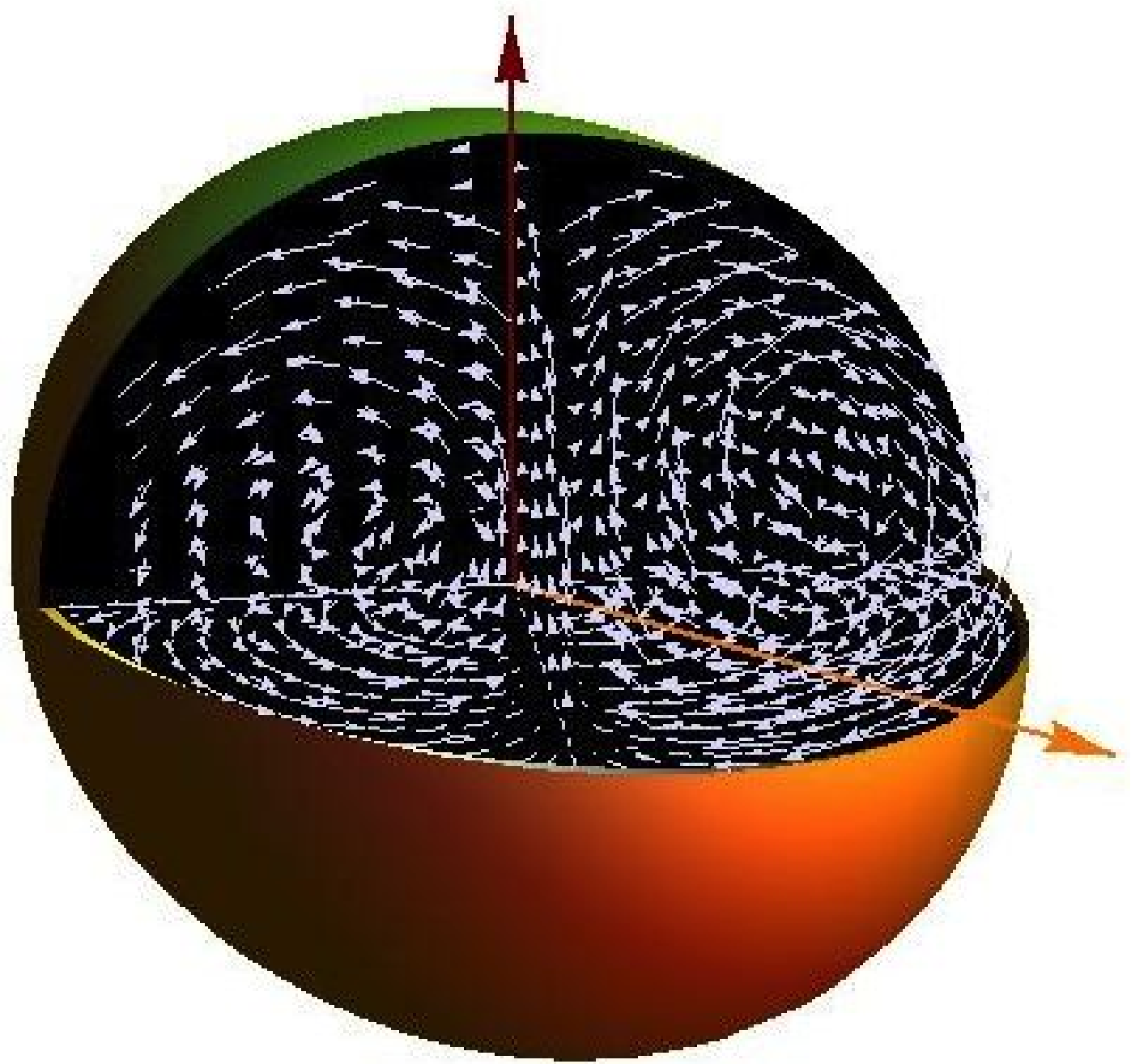}\hskip 70pt
 \includegraphics[width=0.3\textwidth]{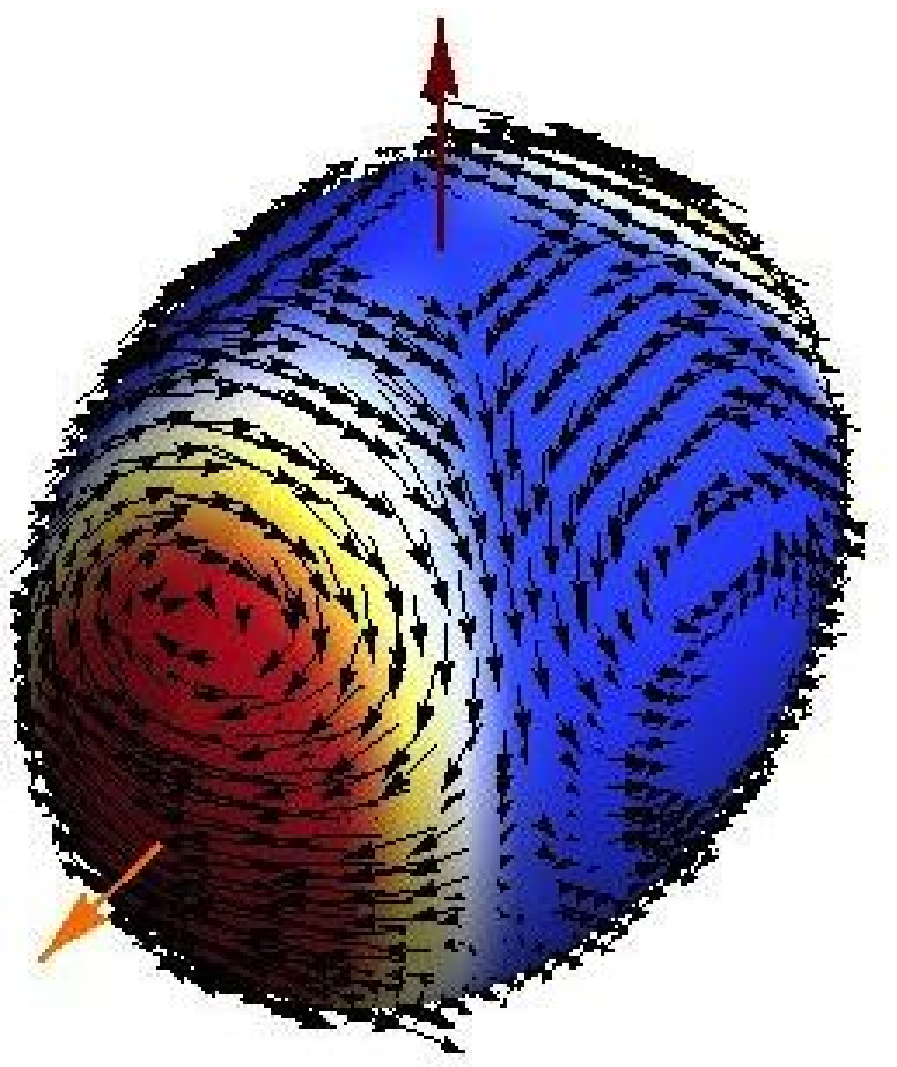}
 \vskip 20pt
 \includegraphics[width=0.375\textwidth]{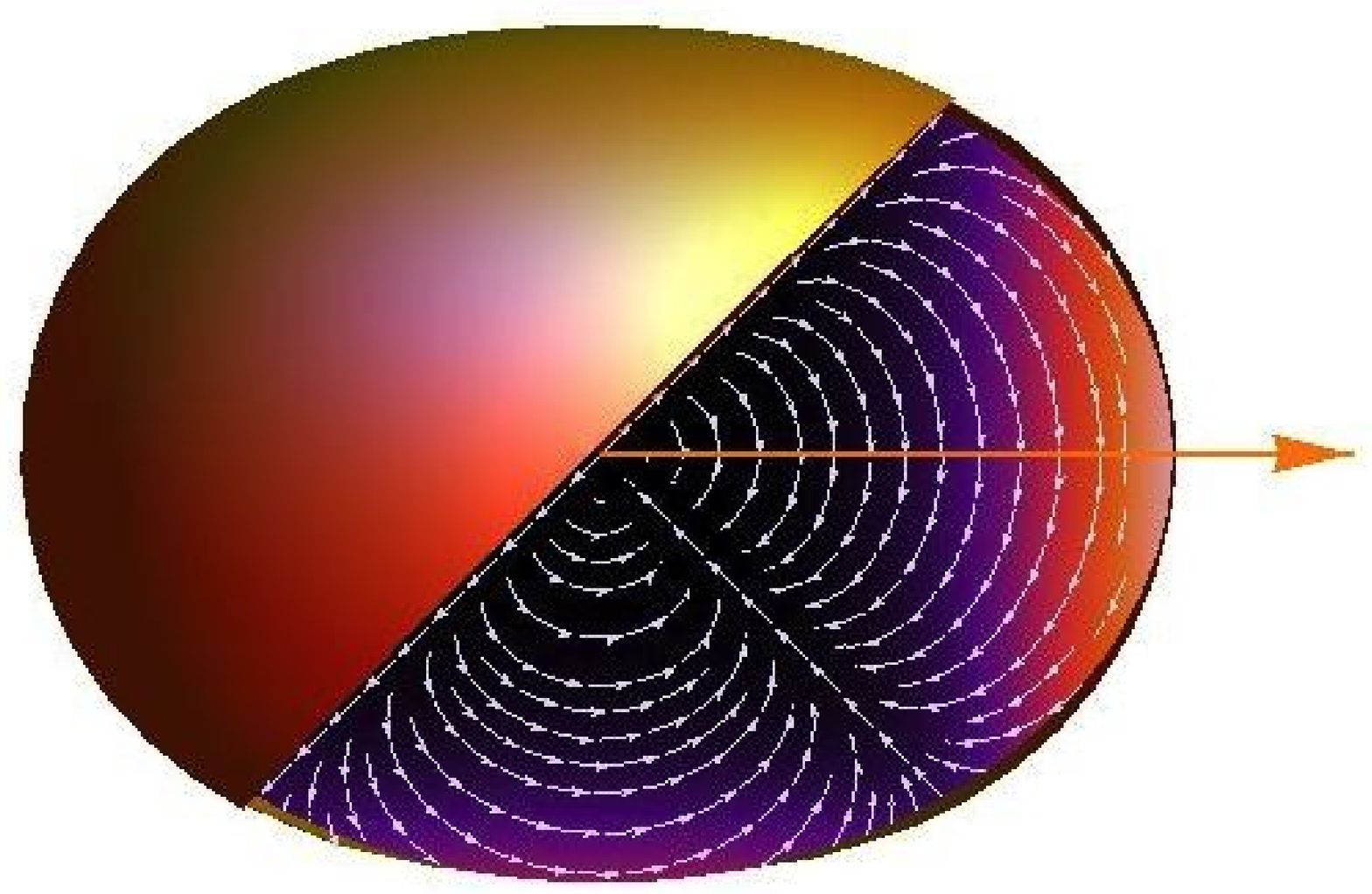}\hskip 70pt
 \includegraphics[width=0.3\textwidth]{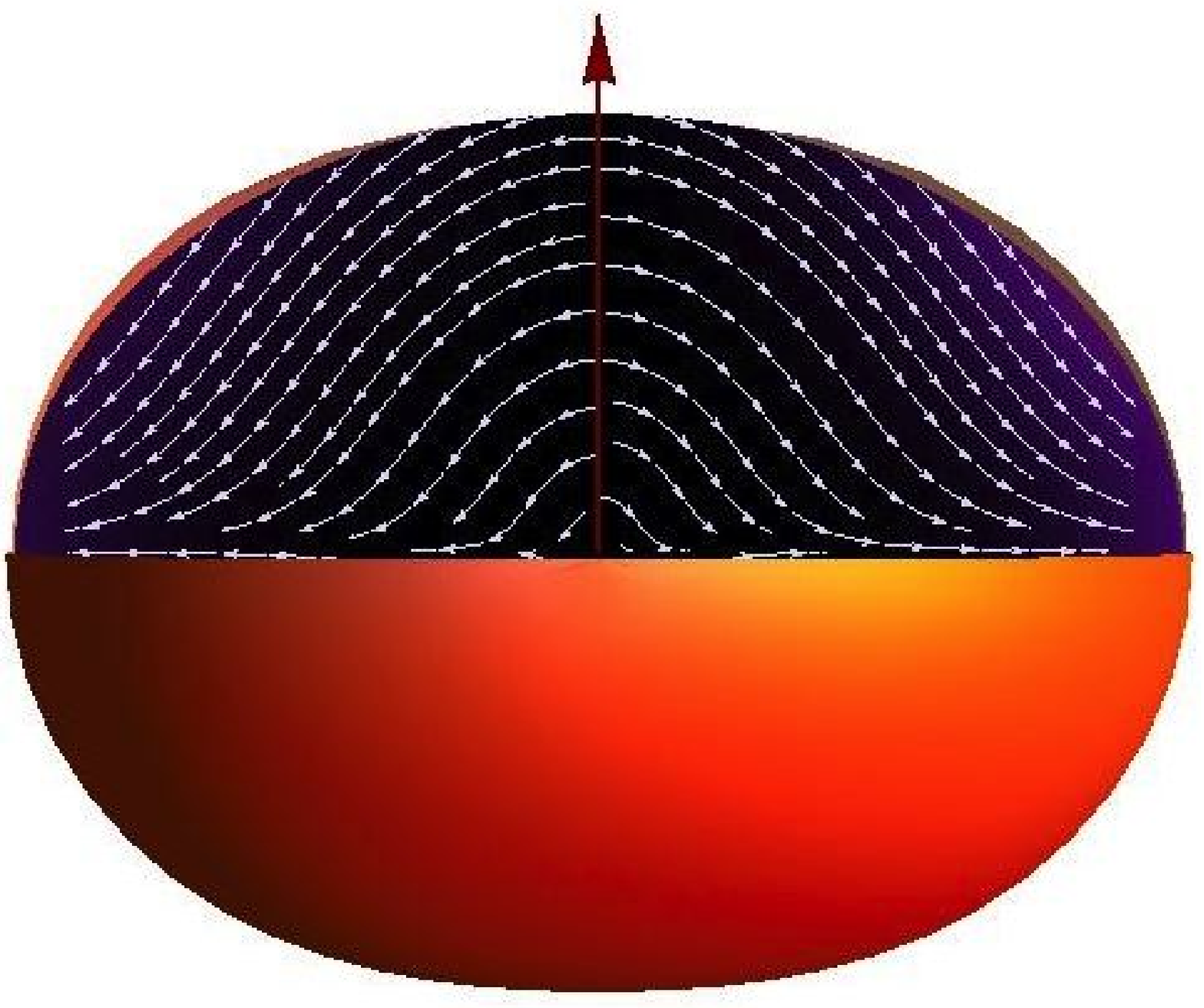} 
  \vskip 20pt
 \end{center}
 \caption{
 {\bf Top-left:} 3--D view of the total (poloidal and toroidal) adiabatic equilibrium tide velocity field (white arrows). 
 The red and orange arrows indicate the direction of the primary's rotation axis and the line of centers respectively. 
 {\bf Top-right:} Representation of this velocity field at the surface of the primary (black arrows); the color-scaled background represents the normalized tidal potential intensity (blue and red for the minimum and maximum values respectively). 
 {\bf Bottom-left:} View of the velocity field (white arrows) in its equatorial plane of symmetry; the color-scaled background represents the velocity value (black and orange for the minimum and maximum values respectively). 
 {\bf Bottom-right:} View of the velocity field (white arrows) in its meridional plane of symmetry; the color-scaled background represents the velocity value (black and purple for the minimum and maximum values respectively).
 \label{FigChamps}
 }
\end{figure*}

\subsection{The dissipative equilibrium tide}

We proceed now with the treatment of the dissipation of the equilibrium tide. We assume that the scale separation (in time and size) between the large scale tidal flow and the turbulent convective motions is sufficient to allow their interaction to be represented by an eddy viscosity $\eta(r)$ acting on the tidal velocity field $\mbf V_\I$. For simplicity, we assume this viscosity to be a scalar function of $r$, although we are aware that it is actually a diagonal tensor, as was shown by Penev at al. (2007).\footnote{Other effects would have to be considered if the rotation were not uniform, as we assumed here; the case of differential rotation will be addressed in a forthcoming paper.} This viscous force causes a lag in the tide and it projects it on what we have introduced above as system (II), giving rise to a flow $\mbf V_\II$, which is in quadrature with the adiabatic tidal flow. For brevity, we shall call it {\it dissipative velocity field}.

\subsubsection{The viscous force}
The treatment of dissipation of the equilibrium tide requires first to express the viscous force induced by the adiabatic velocity field $\mbf F_\II = \mbf F_{\rm visc} \left( \mbf V_\I \right)$. Since $\mbf V_\I$ is solenoidal, that force takes the form:
\begin{equation}
  \mbf F_\II\left(\mbf V_\I\right) =  \nabla\eta  \wedge \left( \nabla \wedge \mbf V_\I \right)
     - \nabla \wedge \nabla \wedge \left( \eta \mbf V_\I \right)
     + \nabla \left( \mbf V_\I \cdot \nabla\eta \right) - \mbf V_\I \nabla^2\eta \; .
\end{equation}

Once again we split the adiabatic velocity field in poloidal and toroidal components:
\begin{equation}
  {\mbf F_\II} \left({\mbf V}_{\I}\right)={\mbf F}_{\II} 
  \left({\mbf V}_{\I}^{P}\right)+{\mbf F}_{\II}\left({\mbf V}_{\I}^{T}\right) \,,
\end{equation}
for which we find the following expressions in terms of the functions $\xi_l^P$ and $\xi_l^T$ we have introduced in (\ref{defxiP}) and (\ref{defxiT}):
\begin{subequations}
\begin{equation}
  \mbf F_{\II}\left(\mbf V_\I^P\right) = \sum_{l}
  \left[ \begin{array}{l}
          f_{l}^{V}\left(x\right) \, \mrm P_2^2 \left(\cos\theta\right) \, \sin\left(2\varphi - ln t\right) \\ \\
          f_{l}^{H}\left(x\right)   \dfrac{\mrm d \mrm P_2^2\left(\cos\theta\right)}{\mrm d\theta}
            \, \sin\left(2\varphi - ln t\right) \\ \\
          2 f_{l}^{H}\left(x\right)  \, \dfrac{\mrm P_2^2\left(\cos\theta\right)}{\sin\theta}
            \, \cos\left(2\varphi - ln t\right)
         \end{array} \right]
   \text{ ,}
\label{FVIP}
\end{equation}
where
\begin{equation}
  \left\{\begin{array}{l}
    f_{l}^{V} = \dfrac{6}{R^2} \, \left\{ \dfrac{\eta}{x^2} \left[ \dfrac{\mrm d^2}{\mrm dx^2} \left(
         x^2 \xi_l^P \right) - 6 \xi_l^P  \right] + 2 \, \dfrac{\mrm d\eta}{\mrm dx} \dfrac{\mrm d \xi_l^P}{\mrm dx} \right\} \\{}\\
   R^2 x f_{l}^{H} \! = \! \eta \left[ \dfrac{\mrm d^3}{\mrm dx^3} \left( x^2\xi_l^P \right) -  
   6 \dfrac{\mrm d}{\mrm dx} \xi_l^P \right]\\
     \hbox{ }\hbox{ }\hbox{ }\hbox{ }\hbox{ }\hbox{ }\hbox{ }\hbox{ }\hbox{ }\hbox{ }\hbox{ }\hbox{ }\hbox{ }\hbox{ }+ \dfrac{\mrm d\eta}{\mrm dx} \left[ x \dfrac{\mrm d^2}{\mrm dx^2} \left( x \xi_l^P \right) +4 \xi_l^P \right]
  \end{array} \right.
\label{ftfthetaAB}
\end{equation}
\end{subequations}
and 
\begin{subequations}
 \begin{equation}
  \mbf F_{\II}\left(\mbf V_\I^T\right) = \sum_{l}
  \left[ \begin{array}{l}
          0 \\ \\
          2 f_{l}^{T}\left(x\right) \, \dfrac{\mrm P_3^2\left(\cos\theta\right)}{\sin\theta}
            \, \sin\left(2\varphi - ln t\right) \\ \\
          f_{l}^{T}\left(x\right)\, \dfrac{\mrm d \mrm P_3^2\left(\cos\theta\right)}{\mrm d\theta}
            \, \cos\left(2\varphi - ln t\right)
         \end{array} \right]
   \text{ ,} \label{FVIT}
 \end{equation}
with
\begin{equation}
f_{l}^{T} = - \frac{1}{R^2}\left\{ \eta \left[ \frac{1}{x}
      \frac{\mrm d^2}{\mrm dx^2} \left( x\, \xi_l^T \right) - 12\frac{\xi_l^T}{x^2}
\right]
      + \frac{\mrm d \eta}{\mrm dx} x
         \frac{\mrm d}{\mrm dx} \left(\frac{\xi_l^T}{x} \right) \right\} \,.
\label{ft}
\end{equation}
\end{subequations}

\subsubsection{Toroidal velocities of the dissipative tide}

Let us rewrite here the momentum equation that governs the dissipative tide  (cf. \ref{eqNS_II}) in the reference frame rotating with the $l$-component of the tidal potential:
\begin{equation}
 \rho_0 \left[  2\mbf\Omega^R\wedge\mbf V_\II \right]
   = -\nabla  P_\II
     - \rho_0 \, \nabla\Phi_\II - \rho_\II \, \nabla\Phi_0 +\mbf F_\II \,.
 \label{eqbisNS_II}
\end{equation}
The viscous force $\mbf F_\II$ has a toroidal component  (\ref{FVIT}) and it generates therefore a toroidal velocity field  $\mbf V_\II^T$ 
which obeys
\begin{equation}
 \rho_0 \left[  2\mbf\Omega^R\wedge\mbf V^T_\II \right]
   = \mbf F_\II \left(\mbf V^T_\I\right) \,.
   \end{equation}
To determine this velocity field, we introduce the stream function $D_{l}\left(x,\theta\right)$ and project $\mbf V^T_\II$ on $\theta$ and $\varphi$:
\begin{equation}
\label{VIIT}
 \mbf V_\II^T = \sum_{l} \left[ \begin{array}{c}
     0 \\ \\
     2  D_{l}\left(x,\theta\right) \, \cos\left(2\varphi-ln t\right) \\ \\
     - \dfrac{\partial}{\partial\theta} \left[ \sin\theta \,  D_{l}\left(x,\theta\right) \right]
        \, \sin\left(2\varphi-ln t\right) \\{}
                \end{array} \right] \;.
\end{equation}
Next, we implement the Coriolis acceleration associated with this velocity field in (\ref{eqbisNS_II}), which we project on the horizontal coordinates:
\begin{subequations}
 \begin{equation}
 2\, \rho_0\, \Omega^R\, \cos\theta \, \frac{\partial}{\partial \theta} \left[ \sin\theta \,  D_{l}\right] = \frac{\partial \mcal F}{\partial\theta} + 2\,f_{l}^{T} \, \frac{\mrm P_3^2\left(\cos\theta\right)}{\sin\theta} \,,
 \end{equation}
 \begin{equation}
 4\, \rho_0\, \Omega^R\, \cos\theta \, D_{l} = \frac{2 \, \mcal F}{\sin\theta} + \,f_{l}^{T}
\, \frac{\mrm d \mrm P_3^2\left(\cos\theta\right)}{\mrm d\theta} \,,
 \end{equation}
where  $\mcal F$ contains the scalar functions of the r.h.s. of the momentum equation other than the toroidal part of the viscous force.
\end{subequations}
As before we proceed with the elimination of  $\mcal F$  and  obtain this expression for $D_{l}$:
\begin{equation}
\label{D}
   D_{l}=  d_{l}\left(x\right) \, \frac{\mrm P_3^2\left(\cos\theta\right)}{\sin\theta}
  \:\text{, where }  
  d_{l} = \frac{3}{\Omega^R \rho_0} \, f_{l}^{T} = \frac{6}{l n \rho_0} \, f_{l}^{T} \,.
\end{equation}

Let us rewrite the toroidal velocity field induced by the dissipation tide:
\begin{equation}
\label{VIITl}
 \mbf V_\II^T = \sum_{l} \left[ \begin{array}{c}
     \\
     0 \\ \\
     \dfrac{6}{\Omega^R \rho_0} \, f_{l}^{T}
          \, \dfrac{\mrm P_3^2\left(\cos\theta\right)}{\sin\theta}
           \, \cos\left(2\varphi-ln t\right) \\ \\
     - \dfrac{3}{\Omega^R \rho_0} \, f_{l}^{T}
         \, \dfrac{\mrm d \mrm P_3^2\left(\cos\theta\right)}{\mrm d\theta}
        \, \sin\left(2\varphi-ln t\right) \\{}
                \end{array} \right] \,.
\end{equation}

The viscous force, as well as its associated toroidal velocity field, produces a mass redistribution inside the star or planet, and thus a perturbation of density $\rho_{\rm II}$ and  gravific potential $\Phi_{\rm II}$. This causes a tidal torque to be applied to the object, which will be responsible for the dynamical evolution of the system. 



\subsubsection{Perturbation of gravific potential $\Phi_\II$}
\label{section_PhiII}

We saw in the previous section that viscous dissipation generates a toroidal velocity field $ \mbf V_\II^T$, which is in quadrature with the toroidal part  $\mbf V_\I^T$ of the adiabatic flow. 
It remains now to determine the gravific potential $\Phi_\II$, which is associated with it, and which is in quadrature with the tidal potential $U$.\\ 

{\it Perturbed Poisson equation}\\

Let us go back to the Navier-Stokes equation (\ref{eqNS_II}) satisfied by the toroidal velocity field $\mbf V_\II^T$ whose expression is given in (\ref{VIIT}), taking into acount relations (\ref{FVIP}) and (\ref{FVIT}) of the viscous forces due respectively to the poloidal and toroidal adiabatic velocity fields.
Projection on $r$ and $\theta$ leads us to the following equations:\\
\begin{subequations}
\label{NSIIproj}
 \begin{multline}
  2\Omega^R \rho_0 \sin\theta \frac{\partial}{\partial\theta} \left(\sin\theta \, D_{l}\right)
    \sin\left(2\varphi-ln t\right) = \\
   -\frac{1}{R}\frac{\partial P_\II}{\partial x} - \rho_0 \frac{1}{R} \frac{\partial \Phi_\II}{\partial x}
   - \rho_\II \frac{1}{R} \frac{\partial \Phi_0}{\partial x}
   + f_{l}^{V} \, \mrm P_2^2\left(\cos\theta\right) \sin\left(2\varphi-ln t\right) \,,
 \end{multline}
and
 \begin{multline}
  2\Omega^R \rho_0 \sin\theta \frac{\partial}{\partial\theta}\left(\sin\theta \, D_l\right)
   \sin\left(2\varphi-ln t\right) = -\frac{1}{R}\frac{1}{x}\frac{\partial P_\II}{\partial\theta}
   - \frac{1}{R} \frac{\rho_0}{x} \frac{\partial \Phi_\II}{\partial\theta} \\
   + \left[ f_{l}^{H} \frac{\mrm d \mrm P_2^2\left(\cos\theta\right)}{\mrm d\theta}
              +2 f_{l}^{T} \frac{\mrm P_3^2\left(\cos\theta\right)}{\sin\theta} \right]
     \sin\left(2\varphi-ln t\right) \,.
 \end{multline}
\end{subequations}

The Coriolis acceleration on the l.h.s. of these equations projects on the associated  Legendre polynomials of second and fourth order $\mrm P_2^2$ and $\mrm P_4^2$, since
\begin{eqnarray}
  2\Omega^R \rho_0 \sin\theta \frac{\partial}{\partial\theta}\left(\sin\theta \, D_l\right) &=&   6 \sin\theta \frac{\partial}{\partial\theta} \mrm P_3^2\left(\cos\theta\right)  f_{l}^{T}  \nonumber \\
  &=& \frac{6}{7}\left[ - 20  \mrm P_2^2  + 6   \mrm P_4^2 \right] f_{l}^{T} \, .
\end{eqnarray}
Thus, the viscous force causes perturbations of density, pressure and internal potential that contain these spherical functions only, and we may expand them as
\begin{equation}
 \label{devscal}
   X_\II = \sum_{l} \left[  X_{2,l}\left(x\right) \, \mrm P_2^2\left(\cos\theta\right)
                     + X_{4,l}\left(x\right) \, \mrm P_4^2\left(\cos\theta\right) \right] \,
                    \sin\left(2\varphi-ln t\right) \,.
\end{equation}
However, we agreed earlier that the tidal potential could be represented by a quadrupole only. In this approximation,  the perturbations in $\mrm P_4^2$ do not contribute to the tidal torque, and we will no longer consider them.
Keeping only the $\mrm P_2^2$ harmonics, dropping the index II,  and setting $s=2$ , we rewrite (\ref{NSIIproj}) as
\begin{subequations}
\label{Ysm}
\begin{empheq}[left=\empheqlbrace\;]{flalign}
-\frac{20}{7} \, 2\Omega^R \rho_0 \,  d_{l} =& -\frac{1}{R}\frac{\mrm d {\widehat P}_{l}}{\mrm dx}
      -\rho_0 \frac{1}{R}\frac{\mrm d {\widehat\Phi}_{l}}{\mrm dx} - g_0 \, {\widehat \rho}_{l} + f_{l}^{V} \,,
      \label{Y22r}\\
\frac{25}{14} \, 2\Omega^R x \rho_0 \, d_{l} =& -{\widehat P}_{l} - \rho_0\,{\widehat \Phi}_{l}
     + R x f_l^H + 5 R x f_l^T \, , \label{Y42theta}
\end{empheq}
\end{subequations}
where we have used the identity $2  \mrm P_3^2 (\cos\theta) = 5 \sin\theta \, \mrm d \mrm P_2^2/ \mrm d \theta$.

By eliminating the  pressure and gravity perturbations, we isolate the term, which will be implemented next in the Poisson equation (\ref{eqPOIS_II}): 
\begin{eqnarray}
\label{PhiRho}
 \quad g_0 \, {\widehat\rho}_{l} &=&
       - \frac{\mrm d\rho_0}{\mrm dx}{\widehat\Phi}_{l} + Z_{l}\left(x\right)\,,\nonumber \\
  \text{where} \quad
   Z_{l} &=& f_l^V - \frac{\mrm d}{\mrm dx} \left(x f_l^H \right)
    + \frac{40}{7} \left[ 3 f_l^T + \frac{\mrm d}{\mrm dx} \left(x f_l^T \right)\right] \,,
\end{eqnarray}
which finally yields
\begin{equation}
  \frac{1}{x} \, \frac{\mrm d^2}{\mrm dx^2} \left( x\,{\widehat\Phi}_{l} \right)
     - \left[ \frac{6}{x^2} - f(x) \right] \, {\widehat\Phi}_{l}
    = -\frac{4\pi \mathcal G R^2}{g_0} \, Z_{l}\,, \label{eqPhiII}
\end{equation}
where as above $f(x) = - (4\pi \mathcal G R / g_0) \, ({\mrm d\rho_0}/{\mrm dx})$. 
It remains to solve that differential equation with the appropriate boundary conditions:   
\begin{equation}
x \, \frac{\mrm d {\widehat\Phi}_l}{\mrm dx} + 3{\widehat\Phi}_l =0 \; {\rm  at} \;  x=1, \,\hbox{and ${\widehat\Phi}_l$ regular at} \; x=0.
\end{equation}

{\it Application to a thin convective envelope}
\vskip 1pc

Equation (\ref{eqPhiII}) allows for an approximate solution in the case where the star, or planet, possesses a convective envelope. If that envelope is not too thick, we may assume that its mass is negligible, so that $g_0(x) \approx g_s/x^2$, and that the function $h$ (cf. \ref{eqh}), may be assumed constant: $h(x)\approx 1$. Then the poloidal and toroidal stream functions take the simple monomial form
\begin{empheq}[left=\quad \empheqlbrace]{flalign}
 \,\xi_l^P = \frac{1}{6} \sigma_l \, \frac{U_{l}}{g_s} \, x^4 \,,\\
 \,\xi_l^T = \frac{2}{5} \sigma_l \, \frac{U_{l}}{g_s} \, x^4 \,,
&&&&&&&&&&&&&&&&&&&&&&
\end{empheq}
where $g_s=g_0(1)$  designates the surface gravity.
From (\ref{ftfthetaAB}), we then deduce that:
\begin{subequations}
\begin{multline}
 \label{f1}
 f_l^V-\frac{\mrm d}{\mrm dx} \left( x f_l^H \right) = 
   - \frac{U_{l}}{R^2 \, g_s} \sigma_l \\
        \times \left(24 x^2\,\eta+24x^3\eta'+4x^4\eta''\right) \,,
\end{multline}
and from   (\ref{ft})   
\begin{multline}
\label{f2}
 3 f_l^T + \frac{\mrm d}{\mrm dx} \left( x f_l^T
\right) =  -\frac{U_{l}}{R^2 \, g_s} \sigma_l \\
  \times \frac{2}{5} \, \left[  48 x^2 \, \eta + 29x^3\,\eta' + 3x^4\,\eta''\right] \,.
\end{multline}
\end{subequations}

Then, introducing these expressions in (\ref{eqPhiII}), we get:
\begin{eqnarray}
  Z_{l} &=&  -\frac{U_l}{R^2 g_s} \, \sigma_l\, \frac{1}{7}\left[ 936\,x^2\,{\eta} + 632\,x^3\,{\eta}' +76 \,x^4\,{\eta}'' \right]\,.
\end{eqnarray}

The last step consists in solving Poisson's equation (\ref{eqPhiII}), which we rewrite
\begin{equation}
 \frac{\mrm d^2}{\mrm dx^2} \left( x{\widehat\Phi}_{l} \right)
     - \frac{6}{x^2}  \, \left(x{\widehat\Phi}_{l}\right)
    = -\frac{4\pi \mathcal G R^2}{g_s} \, x^3 Z_{l} = - z_{l}\left(x\right) \,, \label{eqPhiIIbis}
\end{equation}
letting again $g_0(x) \approx g_s/x^2$ and neglecting the function $f(x)$ compared to $6/x^2$ (for a justification of this approximation we refer to Z66).
This can be done semi-analytically by using the method of Green's integral kernels (Morse \& Feshbach 1953, see also Z66); the formal solution is 
\begin{equation}
{\widehat\Phi}_l(x) = \frac{x^{-3}}{5} \int_0^x \zeta^{3} z_{l}\left(\zeta\right) \, \mrm d\zeta +  \frac{x^2}{5} \int_x^1 \zeta^{-2} z_{l}\left(\zeta\right) \,\mrm d\zeta \, .
\end{equation}
At the surface ${\widehat\Phi}_{\II,l}$, as we name it again to recall the dissipative origin of this tide, takes then the value
\begin{multline}
\label{PhiII}
{\widehat\Phi}_{\II,l}(1) = - \frac{4 \pi}{35} \frac{{\mathcal G} R^2}{g_s} \, \frac{U_l}{R^2 g_s} \, \sigma_l \\
  \times \int_{x_{\rm BCZ}}^1 \left[ 936\,x^8\,{\eta} + 632\,x^9\,{\eta}' +76 \,x^{10}\,{\eta}'' \right] \mrm dx\,,
\end{multline}
where the summation is made over the convective envelope, whose lower boundary is at $x=x_{\rm BCZ}$ and the upper boundary at $x=1$.
Thus the gravific potential due to viscous dissipation ${\widehat\Phi}_{\II,l}$, which is in quadrature with the tidal potential $U_l$, is proportional to the tidal frequency $\sigma_l$, provided that the turbulent viscosity is not a function of $\sigma_l$.

Eq. \eqref{PhiII} can be recast as
\begin{equation}
	\frac{{\widehat \Phi}_{\II,l}(1)}{U_l} = 
		- \, 4 \pi \, \frac{2088}{35} \frac{R^2}{m_A\,g_s} \, \sigma_l \int_{x_{\rm BCZ}}^1 x^8 \eta(x) \, \mrm dx \,.
\label{Phi2surU}
\end{equation}
We see that this ratio, which quantifies the tidal dissipation, is a function of the tidal frequency, unless the turbulent viscosity is inversely proportional to that frequency.
Finally, let us stress that the equilibrium tide dissipation depends strongly on the internal structure of the considered body, and in particular  on the thickness (or the mass) of the convective envelope.


\subsubsection{Turbulent viscosity}
\label{section:visc_dissip}

As stated above, we assumed that the convective motions and the tidal flow are separated enough in temporal and spatial scales  that their interaction can be described by an eddy viscosity $\eta$ acting on the tidal velocity field. As long as this scale separation is fulfilled, we may take $\eta = \rho_0 \nu_t\left(x\right) \approx \rho_0 V_t\left(x\right)  \ell\left(x\right)$, where  $\nu_t$ is the kinematic turbulent viscosity, $V_t$ the vertical rms velocity of the turbulent eddies, and $\ell$ their mean-free path or mixing length. 
The radial profile of $\eta$ is readily drawn from a classical mixing-length model, since
$\eta^3=(3/20) \, \alpha^4 \rho_0^2\,H_P\left(x\right)^3\!F_{\rm conv}\left(x\right)$, where  $H_p=\vert{{\rm d}\ln P_0}/{{\rm d}\ln r}\vert^{-1}$ is the pressure scale-height, $\alpha=\ell /H_P$ the mixing-length parameter and $F_{\rm conv}$ the convective enthalpy flux (cf. Z66, Zahn 1989). 

However, this simple prescription no longer holds when the tidal period $P_{\rm tide}=2 \pi /\sigma_l$ becomes shorter than the life span $t_{\rm conv} = \ell/ V_t$ of the convective elements. One expects then the tidal dissipation to be reduced: just how much is still a matter of debate. On  phenomenological reasons, it seems then natural to replace the mean-free path by the distance traveled by the convective eddies during, say, half a tidal period; then the eddy-viscosity varies proportionally to the tidal period (Z66):
\begin{equation}
\label{nut}
 \nu_t = {1 \over 2} V_t^2 P_{\rm tide} = \pi V_t^2 \sigma_l^{-1}.
\end{equation}
Recently, this scaling has been confirmed through numerical simulations performed by Penev et al. (2007) with a code designed to model stellar convection. They found also that $\nu_t$ is in fact a diagonal tensor, with stronger transport in the vertical than in the horizontal direction, but we shall not take into acount these refinements and simply treat $\eta$ as a scalar.

Another prescription has been proposed by Goldreich \& Keeley (1977), where the turbulent viscosity varies as the square of the tidal period. It implies a much stronger reduction of the viscosity that is incompatible with most observations, for example the circularization of binaries with a red giant component (Verbunt \& Phinney 1995); but it could be adequate for the dynamical tide involving inertial modes, which develop boundary layers that are of much shorter scale than that of the equilibrium tide (cf. Ogilvie \& Lin 2004). 

It thus appears that turbulent dissipation operates in two regimes, depending on how the tidal period compares with the local convective turn-over time, which in a convection zone can vary with depth by several orders of magnitude. 
To ensure a smooth transition between these two regimes, one may take
\begin{equation}
\label{nut2}
 \nu_t = V_t\,\ell \,
    \left[ 1 + \left({2 \, t_{\rm conv} \over P_{\rm tide}} \right)^2  \right]^{-1/2} = V_t\,\ell \,
    \left[ 1 + \left(\frac{t_{\rm conv}\,\sigma_l}{\pi} \right)^2  \right]^{-1/2} \, ,
\end{equation}
as illustrated in Fig.~\ref{Fig_visc}. 

In the upper part of a convective envelope, where the convective turnover time is shorter than the tidal period, $\nu_t$ does not depend on the tidal period; the tidal dissipation varies proportionally to the tidal frequency (cf. Eq.~\ref{Phi2surU}). 
But in the opposite case, when the life span of the convective eddies exceeds the tidal period, which is likely to occur at the base of convection zones, tidal dissipation is independent of the tidal frequency. 
Note that these two regimes still persist once the summation over $\eta$ has been performed in Eq. (\ref{Phi2surU}).

\section{The dynamical evolution of binary systems}

Now that the tidal velocity field and the associated mass redistribution have been obtained, we are ready to determine the tidal torque, which causes the exchange of angular momentum between each component and the orbital motion. 
Let us recall the role played respectively by the adiabatic and the dissipative tide. 
First, the companion exerts a tidal force on the considered component, which is elongated in the direction of the line of centers: this mass redistribution described by the perturbed density $\rho_{\rm I}$ corresponds to the adiabatic tide. 
In the absence of dissipative processes, the induced torque $\Gamma_{\rm I}=\int_{\mathcal{V}} {\partial U}/{\partial \varphi}\,\rho_{\rm I}\,{\rm d}\mathcal{V}$ presents periodic variations of zero average and no secular exchanges of angular momentum are allowed. 
However, turbulent friction 
generates a mass redistribution described by  $\rho_{\rm II}$. 
This dissipative tide produces the torque $\Gamma_{\rm II}=\int_{\mathcal{V}} {\partial U}/{\partial \varphi}\,\rho_{\rm II}\,{\rm d}\mathcal{V}$ of non-zero average, which causes the net exchange of angular momentum between the components of the system as illustrated in Fig. \ref{figMarees}.


\subsection{The dissipation quality factor}
\label{Sect_Q}
The expression of the torque $\Gamma_{\II}$ involves the ratio $\widehat \Phi_{\II,l} / U_l$ which we have established above in Eq. \eqref{Phi2surU}, or equivalently the quality factor $Q$ which we shall now introduce.
For this it is convenient to work in complex notations, with the real part representing the adiabatic tide and the imaginary part characterizing the dissipative contribution.
For the perturbed gravific potential we have then
\begin{equation}
\label{PhiComplex}
	\Phi'\left({\vec r},t\right) = 
	\sum_{l} \Re \left\{ \widehat{\Phi}_{l}(x) P_{2}^{2}(\cos\theta) \exp\left[ {\pmb i} \left( 2\varphi - lnt \right) \right] \right\} \,,
\end{equation}
where we define, following the notations of Eqs. (\ref{XI}-\ref{XII}),
\begin{equation}
	{\widehat \Phi}_l = {\widehat \Phi}_{{\rm I},l} - {\pmb i} \, {\widehat\Phi}_{{\rm II},l}.
\end{equation}
For each Fourier component $l$, the complex second-order Love number ${\tilde k}_2$, which characterizes the deformation response of $A$ to the stress exerted by $B$, is defined by
\begin{equation}
\label{k2Complex1}
	{\tilde k}_2 \left( \sigma_l \right) = \frac{{\widehat \Phi}_l(1)}{U_l} \,.
\end{equation}
Since the turbulent friction introduces a phase delay ${\interval{\epsilon}{-\frac{\pi}{2}}{\frac{\pi}{2}}}$ of the reaction relative to the load, ${\tilde k}_2$ is also of the form
\begin{equation}
\label{k2Complex2}
	{\tilde k}_2 \left( \sigma_l \right) = k_2 \left( \sigma_l \right) \, e^{- \pmb i \epsilon\left( \sigma_l \right) }  \,,
\end{equation}
where $k_2$ is the norm of ${\tilde k}_2$.
From Eq. \eqref{k2Complex1}, we identify the real part of ${\tilde k}_2$ as the adiabatic second-order Love number given by Eq. \eqref{k2def}, thus
\begin{equation}
	k_2^{\rm{ad}} 
		= \Re \left( {\tilde k}_2 \right) 
		= \frac{{\widehat \Phi}_{\I,l}(1)}{U_l} 
		= k_2 \, \cos \epsilon \,.
\end{equation}
The ratio ${{{\widehat \Phi}_{\II,l}(1)}/{U_l} = k_2 \sin \epsilon }$ is similar to ${k_2^{\rm{ad}} = k_2 \cos \epsilon }$, but applied here to the dissipative tide.
By analogy with an electric circuit, we then define the quality factor $Q>0$ (see Ferraz-Mello et al. 2008;  Efroimsky 2012) by
\begin{equation}
\label{Qdef}
	\frac{1}{Q \left( \sigma_l \right)} 
		= \frac{ \left| \Im \; {\tilde k}_2 \left( \sigma_l \right) \right|}{k_2 \left( \sigma_l \right)} 
		= \sin | \epsilon \left( \sigma_l \right) | \,,
\end{equation}
such that the tidal dissipation is given by
\begin{equation}
\label{k2surQ}
\begin{split}
	\frac{k_2\left( \sigma_l \right)}{Q\left( \sigma_l \right)} 
		&= \left| \Im \; {\tilde k}_2 \left( \sigma_l \right) \right|
		=  \left| \frac{{\widehat \Phi}_{\II,l}(1)}{U_l} \right| \\
		&=  4 \pi \, \frac{2088}{35} \frac{R^2}{m_A\,g_s} \times \left| \, \sigma_l \int_{x_{\rm BCZ}}^1 x^8 \eta(x) \, \mrm dx  \, \right| \,,
\end{split}
\end{equation}
where we have recall the expression of Eq. \eqref{Phi2surU}.
Note that the ratio ${k_2\left( \sigma_l \right)}/{Q\left( \sigma_l \right)}$ can be frequency dependent, whereas $k_2^\mrm{ad}$ does not.
This tidal quality factor is also linked to the geometrical lag angle $\delta$, which is half the phase lag $\epsilon$ (see Efroimsky \& Lainey, 2007):
\begin{equation}
\label{k2_eps_delta}
	\frac{k_2\left( \sigma_l \right)}{Q\left( \sigma_l \right)} 
		= k_2 \left( \sigma_l \right) \, \sin | \epsilon \left( \sigma_l \right) |
		= k_2 \left( \sigma_l \right) \, \sin \left| 2 \delta \left( \sigma_l \right) \right| \,.
\end{equation}

\begin{figure}[!htb]
 \centering
 \includegraphics[width=7cm] {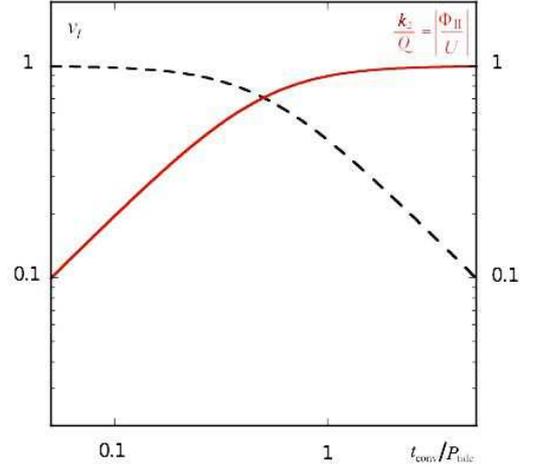}
 \caption{
 The two regimes of turbulent dissipation (Eq.~\ref{nut2}). 
 As long as the local convective turn-over time remains shorter than the tidal period (${t_{\rm conv} < P_{\rm tide}}$), the turbulent viscosity $\nu_t$  (in black dashed line) is independent of the tidal frequency, and the inverse quality factor ${k_2/Q = \left|\Phi_\II / U\right|}$ (in red continuous line), given by Eq. \eqref{k2surQ}, varies proportionally to the tidal frequency $\sigma_l$ (so does also the tidal lag angle). 
When ${t_{\rm conv} > P_{\rm tide}}$, $\nu_t$ varies proportionally to the tidal period, whereas ${k_2/Q = \left|\Phi_\II / U\right|}$ does no longer depend on the tidal frequency. 
Note that $\nu_t$ and ${k_2/Q}$ have been scaled by the value they take respectively for ${t_{\rm conv}/ P_{\rm tide} \rightarrow 0}$ and ${+ \infty}$.
 \label{Fig_visc}
 }
\end{figure}

Recalling {Eq.~\eqref{Phi2surU}} of ${{{\widehat \Phi}_{\II,l}(1)}/{U_l}}$, we notice that when the turbulent viscosity does not depend on tidal frequency, the tidal lag angle ${\delta \left(\sigma_l\right) = \sigma_l \, \Delta t \left( \sigma_l \right)}$ is proportional to the tidal frequency, which means that the time lag ${\Delta t\left( \sigma_l \right)}$ of the tide is then constant, and takes the same value for each Fourier component of the tide: this is what has been called the weak friction approximation (Hut 1981).
It corresponds to the slow tide (${t_\mrm{conv} < P_\mrm{tide}}$), as illustrated in Fig. \ref{Fig_visc}.
In the opposite case where ${t_\mrm{conv} > P_\mrm{tide}}$, the tidal angle $\delta$ is independent of the tidal frequency, and so are also the phase lag $\epsilon$ and the quality factor $Q$, such that the time lag ${\Delta t = \delta / \sigma_l}$ is inversely proportional to the tidal frequency.

\subsection{The disturbing function}

Here, we shall use the disturbing function method as described in Brouwer \& Clemence (1961), Kaula (1962), Yoder (1995-97), Correia \& Laskar (2003a,b) and Mathis \& Le Poncin-Lafitte (2009) (hereafter MLP09) to determine $\Gamma_{\rm II}$ and the associated variation of the bodies' angular momentum and of the orbital keplerian elements of B (namely, $a$ and $e$).\\

\begin{figure}[!htb]
\begin{center}
\includegraphics[width=\linewidth]{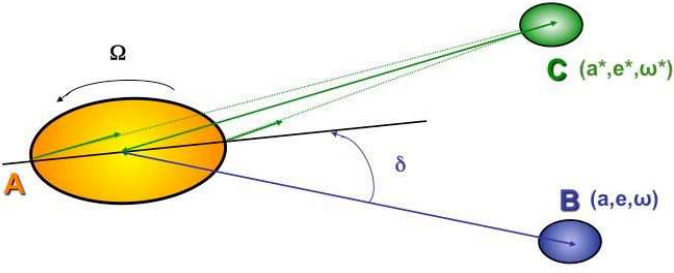}
\vskip 15pt
\caption{
Here, we illustrate the used method of the perturbing function. 
The primary A is deformed because of the tides exerted by the perturber B. 
Because of A's internal friction processes, the tidal bulge is shifted from the line of centers with the tidal angle $\delta_l$ given in Eq. (\ref{k2surQ}) for a given $l$ Fourier component of the tidal potential. 
The keplerian elements of B's orbit are $a$, $e$, and $M$. 
Next, a third body C, which orbits around A with a mean motion $n^{*}$, is introduced, and the variations of the keplerian elements of its orbit ($a^{*},e^{*},M^{*}$) are derived. 
After using Lagrange's equation (Eqs. \ref{evola}-\ref{evole}), we finally consider that the orbiter and the perturber are the same ({\it i.e.} C=B).
\label{figFunPert}
}
\end{center}
\end{figure}

The first step to achieve this goal is to introduce a supplementary body C with a mass $m_{\rm C}$, which we call the orbiter, to the primary A and the perturber B (see Fig. \ref{figFunPert} and Kaula 1962). 
The keplerian elements of its orbit are $a^{*}$, $e^{*}$, and $M^{*}=n^*t$ with $n^*$ the associated mean motion.

Next, we identify the disturbing function on C to the perturbed external gravific potential of A:
\begin{equation}
	\mcal R
	=\Phi'\left({\vec r}_{\rm C},t\right)
	=\sum_{l^{*}}\Re\left\{\widehat{\Phi}_{l^{*}}\left(x_{\rm C}\right)P_{2}^{2}\left(0\right)\exp\left[{\pmb i}\left(2\varphi_{\rm C}-l^{*}n^{*}t\right)\right]\right\},
\label{Cdef}
\end{equation}
where ${\vec r}_{\rm C}\equiv\left(r_{\rm C}=x_{\rm C} R, \theta_{\rm C}=\pi/2, \varphi_{\rm C}\right)$, with $x_{\rm C}>1$, is the position of C in the equatorial plane of A, which is also the plane of the orbit, and where the index $l^{*}$ is associated to the Fourier expansion of the orbital motion of C.\\

From now on, we choose to follow the systematic method developed in MLP09 to derive the expression of ${\mathcal R}$ as a function of the keplerian elements of B and C. 
In this work devoted to the study of the tidal dynamics of extended bodies, the perturbing function is derived for components X having an external gravific potential written as
\begin{equation}
{\Phi}^{\rm X}\left(\vec r,t\right)=\mcal G\sum_{s_{\rm X}=0}^{\infty}\sum_{m_{\rm X}=-s_{\rm X}}^{s_{\rm X}}M_{s_{\rm X},m_{\rm X}}\frac{Y_{s_{\rm X}}^{m_{\rm X}}\left({\widetilde\theta}_{\rm X},{\widetilde\varphi}_{\rm X}\right)}{{\widetilde r}_{\rm X}^{\,s_{\rm X}+1}}\,,
\end{equation}
where $\left({\widetilde r}_{\rm X}, {\widetilde \theta}_{\rm X}, {\widetilde\varphi}_{\rm X}\right)$ are the spherical coordinates attached to their equatorial planes. We introduce the gravitational moments of X
\begin{equation}
M_{s_{\rm X},m_{\rm X}}=\frac{4\pi}{2s_{\rm X}+1}\int_{V_{\rm X}}{\widetilde r}_{\rm X}^{\,s_{\rm X}}Y_{s_{\rm X}}^{m_{\rm X}}\left({\widetilde\theta}_{\rm X},{\widetilde\varphi}_{\rm X}\right)\rho_{\rm X}{\rm d}V_{\rm X}\,,
\end{equation}
where ${\rm d}V_{\rm X}={\widetilde r}_{\rm X}^2{\rm d}{\widetilde r}_{\rm X}\sin{\widetilde\theta}_{\rm X}\,{\rm d}{\widetilde\theta}_{\rm X}{\rm d}{\widetilde\varphi}_{\rm X}$; the non-spherical components of ${\Phi}^{\rm X}$ are due to tides and to the own structural shape of the body. 
We use spherical harmonics
\begin{equation}
Y_{s}^{m}\left(\theta,\varphi\right)=\left(-1\right)^{\frac{\vert m\vert+m}{2}}\sqrt{\frac{2 s+1}{4\pi}\frac{\left(s-\vert m\vert\right)!}{\left(s+\vert m\vert\right)!}}P_{s}^{m}\left(\cos \theta\right)\exp\left[\pmb i\, m\varphi\right]\,.
\end{equation}
Here, we apply this general formalism to the interaction between A, which has a tidal gravitational moment $M^{\rm A}_{2,2}$ (the quadrupolar approximation is assumed for the perturbed external gravific potential of A as for the tidal potential $U$), and C, which is assumed to be ponctual as well as B. 
Then, the Kaula's expansion (Kaula 1962) allows to convert the expansion on spherical harmonics on to a Fourier one on keplerian elements (see Eq. 79 in MLP09). 
Applied to our coplanar case, this gives for ${\rm X}\equiv\left\{{\rm B},{\rm C}\right\}$
\begin{equation}
\frac{Y_{2}^{2}\left(\pi/2,\varphi_{\rm X}\right)}{r_{\rm X}^{3}}=\frac{1}{4}\sqrt{\frac{15}{2\pi}}\frac{1}{a_{\rm X}^{3}}\sum_{l_{\rm X}}G_{2,0,l_{\rm X}-2}\left(e_{\rm X}\right)\exp\left[\pmb i\,\Psi_{l_{\rm X}}\right]\,,
\end{equation}
where
\begin{equation}
\Psi_{l_{\rm X}}=l_{\rm X}\,M_{\rm X}+2\,\omega_{\rm X}-2\,\widetilde\Theta_{\rm A}\,.
\end{equation}
The spherical coordinates of the X position in the A reference frame are $\left(r_{\rm X},\pi/2,\varphi_{\rm X}\right)$ while $l_{\rm X}\equiv\left\{l,l^*\right\}$, $a_{\rm X}\equiv\left\{a,a^{*}\right\}$, $e_{\rm X}\equiv\left\{e,e^{*}\right\}$, and $M_{\rm X}=n_{\rm X}t\equiv\left\{M=n t,M^{*}={n^*}t\right\}$. 
The argument of the periapsis $\omega_{\rm X}\equiv\left\{0,\omega^{*}\right\}$ is also introduced since it could be different for B and C. 
Finally, $\widetilde\Theta_{\rm A}$ is the mean sideral angle between the minimal axis of inertia and the (A${\vec {\rm X}}$) axis, and thus $\Omega ={\rm d}\widetilde\Theta_{\rm A}/{\rm d}t$.\\

This transformation leads in the general case to the expansion of ${\mathcal R}$ given in Eq. (144) in MLP09, which reduces to the following expression, once applied to our case:
\begin{equation}
\label{Rdef}
{\mathcal R} \!=\! - \sum_{l,l^{*}} \! \Re \left\{
			k_{2}^\mrm{ad} \, {\widetilde{\mathcal Z}}_{l} \left(\sigma_{l}\right) \! \frac{4\pi}{5} \frac{ \mcal G m_{B} R^5 }{a^3{a^*}^3} \mcal H_{l} (e) \mcal H_{l^{*}} \! \left(e^{*}\right) 
			e^{\pmb i \left( \Psi_{l^*} - \Psi_{l} \right)} 
		\right\}
\end{equation}

where ${\widetilde{\mathcal Z}}_l \left( \sigma_l \right)$ is the so-called dissipative impedance, defined as
\begin{equation}
	{\widetilde{\mathcal Z}}_l  \left( \sigma_l \right) = \frac{ {\tilde k}_2  \left( \sigma_l \right) }{ k_2^\mrm{ad}  \left( \sigma_l \right) }
														= 1 - \pmb i \frac{ {\widehat\Phi}_{\II,l}(1) }{ {\widehat\Phi}_{\I,l}(1) } 
														= 1 - \pmb i \, \tan \epsilon \left( \sigma_l \right) \:,
\end{equation}
while
\begin{equation}
\mcal H_{l_{\rm X}}\left(e_{\rm X}\right)=\left(\frac{15}{32\,\pi}\right)^{1/2}G_{2,0,l_{\rm X}-2}\left(e_{\rm X}\right)\,.
\end{equation}

Thus, we can recast Eq. \eqref{Rdef} as
\begin{equation}
	{\mathcal R}\!=\!-\sum_{l,l^{*}}\! \Re \left\{ \tilde{k}_{2} \left(\sigma_{l}\right) \! \frac{4\pi}{5}\frac{\mcal G m_{B} R^5}{a^3{a^*}^3}\mcal H_{l}\left(e\right)\mcal H_{l^{*}}\!\left(e^{*}\right) e^{\pmb i\left(\Psi_{l^*}-\Psi_{l}\right)} \right\} \,.
\end{equation}

The adiabatic component ${\mathcal R_\I}$ (which only involves ${\Re \left(\tilde{k}_2 \right)}$) of the perturbing function ${\mathcal R}$ contributes only as zero-average perturbations to the angular momentum and the orbital elements, whereas its dissipative component ${\mathcal R_\II}$ ({\it i.e.} the one proportional to ${\Im \left(\tilde{k}_2 \right)}$) given by
\begin{equation}
\begin{split}
	{\mathcal R}_{\II} 
		&= - \sum_{l,l^{*}} \! \left\{ 
			\frac{ k_2 \left(\sigma_l\right) }{ Q \left(\sigma_l\right) } \frac{4\pi}{5} \frac{ \mcal{G} m_{B} R^5 }{ a^3 {a^*}^3 } \, \mcal{H}_l(e) \, \mcal{H}_{l^*} \! \left( e^* \right) \sin \left( \Psi_{l^*}-\Psi_{l} \right) \right\}    \\
		&= - \sum_{l,l^{*}} \! \left\{ 
			\frac{ {\widehat\Phi}_{\II,l}(1) }{ U_l } \frac{4\pi}{5} \frac{ \mcal{G} m_{B} R^5 }{ a^3 {a^*}^3 } \, \mcal{H}_l(e) \, \mcal{H}_{l^*} \! \left(e^*\right) \sin \left( \Psi_{l^*}-\Psi_l \right) \right\}	
\end{split}
\label{RII}
\end{equation}
acts on secular time-scales.

We thus obtain for the mean disturbing function the same form as Correia \& Laskar (2003a). 
The main difference is that in our case we link the viscous dissipation with the physical processes operating in the considered bodies,i.e. here turbulent convection. 
Since the turbulent viscosity depends on the tidal frequency, we have to treat the dissipation separately for each mode $l$ of the tidal potential, as it has been already shown in Zahn (1977) and Ogilvie \& Lin (2007). 
The general method presented here, based on Zahn (1977) and MLP09, thus allows to go beyond the usual constant $Q$ or tidal lag approximations that can only constitute a first approach as shown in \S \ref{section:visc_dissip} and in Ogilvie \& Lin (2004). 
In a near future, we will be able to apply this method to the fluid dynamical tide and to the anelastic tide for planets that have a rocky part (Efroimsky \& Lainey 2007; Henning et al. 2009).

\subsection{Equations of dynamic evolution of the system}
We are now ready to derive the evolution equations for the angular momentum of A and for the orbital elements of C. First, following Yoder (1995-1997) and Correia \& Laskar (2003a,b), we obtain the equation for the angular momentum of the primary
\begin{equation}
\frac{{\rm d}L_{\rm A}}{{\rm d}t}= \frac{{\rm d} \, I_{A}\Omega}{{\rm d}t}=m_C \partial_{\widetilde \Theta_{\rm A}}{\mathcal R}_{\II},
\label{evolO}
\end{equation}
where $I_{A}$ designates its inertial momentum. Next, applying Lagrange's equations (see for example Brouwer \& Clemence 1961), we get those for the evolution of the semi-major axis of the C's orbit and its eccentricity:
\begin{equation}
\frac{{\rm d}a^{*}}{{\rm d}t}=\frac{2}{n^{*}a^{*}} \partial_{M^{*}}\mcal R_{\II} \,,
\label{evola}
\end{equation}
\begin{equation}
\frac{{\rm d}e^{*}}{{\rm d}t}=-\frac{\sqrt{1-{e^*}^{2}}}{n^{*} \, {a^*}^{2}{e^*}}\partial_{\varpi^{*}}\mcal R_{\II} +
\frac{1-{e^*}^{2}}{n^{*}\,{a^*}^{2}{e^*}}\partial_{M^{*}} \mcal R_{\II} \,.
\label{evole}
\end{equation}
Finally, variations of its mean anomaly $M^{*}$ given by
\begin{equation}
\frac{{\rm d}M^{*}}{{\rm d}t} = n^{*}-\frac{2}{n^{*}\,a^{*}}\partial_{a^{*}}\mcal R_{\II} -
\frac{1-{e^*}^{2}}{{n^*}\,{a^*}^{2}e^{*}}\partial_{e^{*}}\mcal R_{\II}
\end{equation}
are filtered.\\

Applying previous Eqs. (\ref{evola}-\ref{evole}) to ${\mathcal R}_{\rm II}$ (Eq. \ref{RII}) using results derived in MLP09 (Eqs. 134-136-137), we get assuming that the orbiter is the perturber ({\it i.e.} B=C)

\begin{equation}
 \frac{\mrm d L_A}{\mrm dt} = - \frac{8\pi}{5} \frac{\mcal G m_{B}^{2} R^5}{a^6} \sum_l \left\{ \frac{{\widehat\Phi}_{\II,l}(1)}{U_l} \left[\mcal H_{l}\left(e\right)\right]^2 \right\} 
 							= \Gamma_{\II}\,,
 \label{D1}
\end{equation}

where we identify $\Gamma_{\II}$ the secular torque applied to A, and

\begin{equation}
	\frac{1}{a} \, \frac{\mrm da}{\mrm dt} = - \frac{2}{n} \frac{4\pi}{5} \frac{\mcal G m_{B} R^5}{a^8} \sum_l \left\{ l\frac{ {\widehat\Phi}_{\II,l}(1) }{U_l} \left[ \mcal{H}_{l}(e) \right]^2 \right\} \,,
 \label{D2}
\end{equation}

\begin{eqnarray}
	\lefteqn{ \frac{1}{e} \, \frac{\mrm de}{\mrm dt} = - \frac{1}{n} \frac{1-e^2}{e^2} \frac{4\pi}{5} \frac{\mcal G m_{B} R^5}{a^8} \times } \nonumber\\
	& &\sum_l \left\{ \left[ 2 \left( 1 - \frac{1}{\sqrt{1-e^2}} \right) + (l-2) \right] \frac{ {\widehat\Phi}_{\II,l}(1) }{U_l} \left[ \mcal{H}_{l}(e) \right]^2 \right\} \,,
 \label{D3}
\end{eqnarray}

where we recall that (Eqs. \ref{Phi2surU}-\ref{k2surQ}):

\begin{equation}
	\left|\frac{ {\widehat\Phi}_{\II,l}(1) }{U_l}\right| = \frac{k_2}{Q \left(\sigma_l\right)}
											= 4 \pi \frac{2088}{35} \frac{R^2}{m_A\,g_s} \left|\,\sigma_l \int_{x_{\rm BCZ}}^{1}  x^8 \, \eta (x,\sigma_l) \, {\rm d}x\,\right|.
\end{equation}
From these equations one readily draws the characteristic synchronisation and circularisation times: 
\begin{equation}
	\frac{1}{t_{\rm sync}}=-\frac{1}{\left(\Omega-n\right)}\frac{1}{I_{A}}\frac{\mrm d L_A}{\mrm dt}\hbox{ }\hbox{ }\hbox{and}\hbox{ }\hbox{ }\frac{1}{t_{\rm circ}}=-\frac{1}{e}\frac{{\rm d}e}{{\rm d}t}.
\end{equation}

Let us now comment this system of equations. First, as it has been already discussed in the previous section, the equilibrium tide dissipation is consistently described with its dependence on the tidal frequency. 
Moreover, we never made any assumption concerning the value of the eccentricity. More precisely, we chose to keep the most general expansion in $e$ of the tidal potential $U$ and the resulting disturbing function $\mcal R_{\II}$. This means that, provided  we take a large enough number of modes in the tidal potential expansion, as illustrated in Savonije (2008), we are able to treat the general case of eccentric orbits. Therefore, the present formalism allows to take into account the complexity coming both from a general eccentric orbital motion and from the physical processes that act on the tidal velocity field it generates. 

Finally, let us remark that the equilibrium states resulting from Eqs. (\ref{D1}-\ref{D2}-\ref{D3}) will differ from those predicted assuming a constant $Q$ factor or tidal lag because of the different dissipation of each Fourier component of the velocity field; this will be explored in paper II where inclination and obliquity will be taken into account.

\section{Conclusion}

{The originality of this work is the formulation of the equilibrium tide problem in the rotating frame associated with each Fourier component of the tidal potential. 
This allows to filter correctly the dynamical tide (due to the free oscillations of the considered body), and to avoid the so-called `pseudo-resonances' which appeared in Z66 when the equations were written in an inertial frame.
Morever, the solenoidal behavior of the velocity field of the equilibrium tide, which was questioned by Scharlemann (1981), is confirmed. 
We revisit the derivation of the adiabatic and the dissipative components of the equilibrium tide (the velocity fields and the associated perturbations of the gravific potential) with taking into account the most general expansion of the tidal potential in the case where the orbital spin and those of each body are aligned. 
This allows to treat the general eccentric case by taking a sufficient number of Fourier components into account. 
We obtain the associated rate of dissipation of the tidal kinetic energy into heat, given by the $k_2/Q$ ratio, which depends on the thickness of the convective envelope and on the tidal frequency $\left(\sigma_l\right)$. 
Indeed, because of the dependance of the turbulent friction on $\sigma_l$, two dissipation regimes are identified. 
First, as long as the local convective turn-over time remains shorter than the tidal period ($t_{\rm conv} < P_{\rm tide}$), the turbulent viscosity $\nu_t$  is independent of $\sigma_l$, and the inverse quality factor $k_2/Q$ varies proportionally to the tidal frequency (so does also the tidal lag angle). 
When $t_{\rm conv} > P_{\rm tide}$, $\nu_t$ varies proportionally to the tidal period $\left(\propto \sigma_l^-1\right)$, whereas $k_2/Q$ does no longer depend on the tidal frequency. 
This emphasizes the need of going beyond the classical constant tidal lag approximation, not only for the dynamical tide, as it was shown Ogilvie \& Lin (2004), but also for the equilibrium tide. 

In the forthcoming paper, we shall take into account the orbital inclination and the obliquities of the bodies. 
Moreover, we will examine the impact of differential rotation on the equilibrium tide, and that of an anelastic core, which may well be present in giant planets. We plan also to check theoretical predictions againt observations in our Solar system as well as in exoplanetary systems and binary stars.}

\section*{Acknowledgments}

The authors express their gratitude to the referee M. Efroimsky for his constructive remarks and suggestions. 
They
thank V. Lainey, C. Le Poncin-Lafitte,  A.-S. Brun,  S. Udry, R. Mardling and A. Triaud for fruitful discussions during this work, which was supported in part by the Programme National de Plan\'etologie (CNRS/INSU), the EMERGENCE-UPMC project EME0911, and the CNRS {\it Physique th\'eorique et ses interfaces} program.


\end{document}